\documentclass[final,5p,times,twocolumn,authoryear]{elsarticle}

\usepackage{amssymb}
\usepackage{amsmath}
\usepackage{xcolor}

\definecolor{linkcolor}{HTML}{1A5FB4}
\usepackage[colorlinks=true,
            allcolors=linkcolor,
            breaklinks=true,
            pdftitle={Real-space overlap is not enough: ambiguity in nanobeam
                      iterative ptychography},
            pdfauthor={S. M. Ribet, M. Danaie, W. P. M. de Kleijne,
                       A. R. C. McCray, F. Allars, C. S. Allen, C. Ophus,
                       G. Varnavides}]{hyperref}
\usepackage{cleveref}
\crefname{figure}{Fig.}{Figs.}
\Crefname{figure}{Fig.}{Figs.}
\crefname{equation}{Eq.}{Eqs.}
\Crefname{equation}{Eq.}{Eqs.}
\crefname{table}{Table}{Tables}
\Crefname{table}{Table}{Tables}
\crefname{section}{Section}{Sections}
\Crefname{section}{Section}{Sections}

\newcommand{\eA}{e$^-$/\AA$^2$}

\emergencystretch=\maxdimen
\begin{document}

\begin{frontmatter}

\title{Real-space overlap is not enough: ambiguity in nanobeam iterative ptychography}

\author[berkeley]{Stephanie M. Ribet\fnref{equal}}
\ead{sribet@lbl.gov}

\affiliation[berkeley]{
organization={National Center for Electron Microscopy, Molecular Foundry, Lawrence Berkeley National Laboratory},
city={Berkeley},
postcode={94720},
state={CA},
country={United States of America}
}

\author[epsic]{Mohsen Danaie\fnref{equal}}

\affiliation[epsic]{
organization={Electron Physical Science Imaging Centre, Diamond Light Source},
addressline={Harwell Science and Innovation Campus},
city={Didcot},
postcode={OX11 0DE},
state={Oxfordshire},
country={United Kingdom}
}
\ead{mohsen.danaie@diamond.ac.uk}
\fntext[equal]{These authors contributed equally to this work.}

\author[delft]{Willem P.M. de Kleijne\fnref{equal}}
\affiliation[delft]{
    organization={Department of Imaging Physics, Delft University of Technology},
    addressline={Lorentzweg 1},
    city={Delft},
    postcode={2628 CJ},
    country={The Netherlands}
}

\author[stanford]{Arthur R. C. McCray}
\affiliation[stanford]{
    organization={Department of Materials Science and Engineering, Stanford University},
    city={Stanford},
    postcode={94305},
    state={CA},
    country={United States of America}
}

\author[epsic]{Frederick Allars}

\author[epsic]{Christopher S. Allen}

\author[stanford]{Colin Ophus}

\author[delft]{Georgios Varnavides}
\ead{g.varnavides@tudelft.nl}

\begin{abstract}

High-resolution iterative ptychography typically relies on a well-aligned, high-convergence-angle electron probe.
Here we explore whether it can instead be performed at small convergence angles, relaxing the need for probe correctors and enabling experiments at low accelerating voltages or with a de-excited objective lens, as in Lorentz mode.
Through experiments and simulations, we show that once the convergence angle is small enough that no diffracted disks overlap, the resulting reconstruction is ambiguous, posing a significant challenge for robust interpretation of results.
This challenge arises because of the lack of interference between Bragg disks in the recorded diffraction pattern intensity, leading to no phase information for each reflection.
Reconstructions with these data lead to degenerate objects in which rigid translations of the lattice and reversals of contrast of the object produce the same error between experimental data and the ptychography forward model. 
Increasing the real-space overlap between probe positions does not lift this degeneracy.
An amorphous substrate can supply the missing phase relationships by giving the Bragg beams support in the gaps between disks.
This phasing is fragile, however, and survives only where the forward model matches the experiment.
At fixed dose, either constraining the object to be a pure phase object or introducing thermal motion into the forward model is enough on its own to make the solution non-unique, highlighting why our experimental reconstructions below the overlap threshold are ambiguous despite ample dose and real-space redundancy.
Most troublingly, the lattice spacing and orientation are always recovered correctly, so a non-unique reconstruction looks convincing and can be diagnosed only by repeating the reconstruction from different starting points.

\end{abstract}


\end{frontmatter}

\section{Introduction}

\begin{figure*}[t]
\centering
\includegraphics[width = \textwidth]{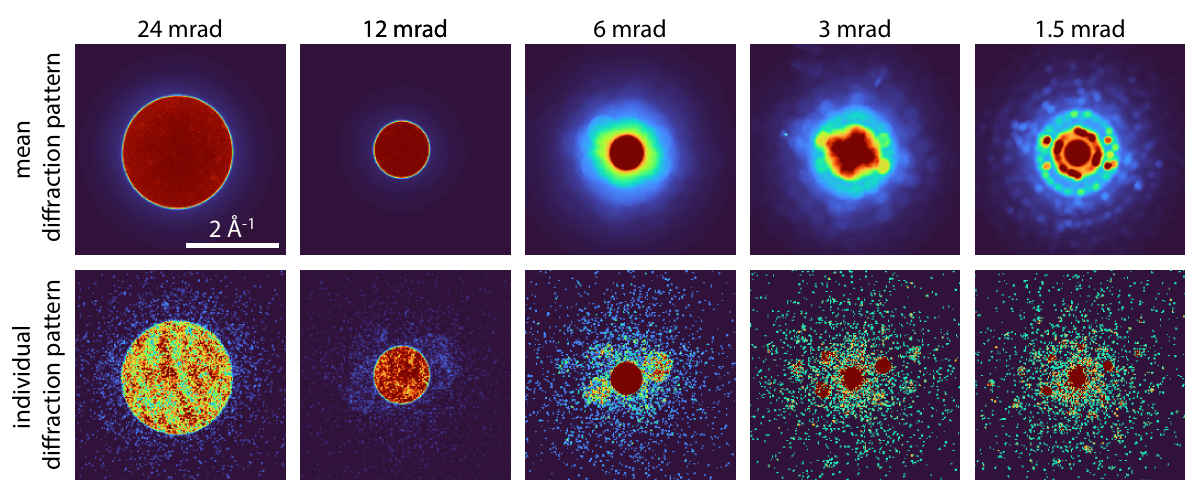}
  \caption{
    Mean and individual diffraction patterns from 4D-STEM datasets acquired from the same field of view with different convergence angles.
    Large convergence angle data have significant overlap and individual patterns show shadow images typical of high-resolution ptychography experiments.
    Small convergence angle data have large separation of disks and a higher signal-to-noise ratio typical of nanobeam diffraction experiments.
  }
  \label{fig:dps}
\end{figure*}

Four-dimensional scanning transmission electron microscopy (4D-STEM), in which a full diffraction pattern is recorded at every probe position in a STEM experiment, has given rise to a family of phase retrieval techniques that computationally recover the phase of the specimen transmission function~\citep{Ophus_2019, Ophus_2023}.
By leveraging the position-sensitive scattering information contained in the diffraction plane, these techniques offer substantial advantages over the widely used annular dark-field (ADF) imaging, most notably improved electron dose efficiency and a linear contrast transfer function~\citep{varnavides2026abcs, varnavides2026beyond}.
Ptychography is among the most widely used of these approaches, and is steadily becoming a core part of modern STEM practice~\citep{jiang2018electron, Chen_2020, pelz2023solving}.
The term ``ptychography'' spans two distinct classes of algorithm, which differ in how they invert the measured data.

Direct methods, including single-sideband (SSB) ptychography~\citep{Rodenburg_1993, Yang_2016} and Wigner distribution deconvolution (WDD)~\citep{Rodenburg_1992, Li_2014, Yang_2017}, recover the object phase in a single step by deconvolving the effect of the probe from the recorded intensities.
Under the weak phase object approximation (WPOA), direct ptychography techniques such as SSB, WDD, optimum bright-field (OBF), parallax imaging and tilt-corrected bright-field (tcBF) STEM all follow from the same linear forward model~\citep{Spoth_2017, Yu_2025, Varnavides_2023, Varnavides_2024, Ooe_2021, Ooe_2024, varnavides2026abcs}.
They differ only in how they weight the bright-field intensities, how they normalize the resulting coherent sum, and how much of the aperture-overlap kernel they retain.
Because they only require a single pass over the data, direct methods are fast enough to support streaming reconstructions during acquisition~\citep{Yu_2022, Pelz_2022, Strauch_2021, bekkevold2024ultra}, and their dose efficiency makes them well suited to beam-sensitive specimens~\citep{Yu_2025, Ooe_2024}.

Iterative ptychography, by contrast, is a more complex and computationally demanding approach.
Leveraging a model of beam-specimen interactions, the algorithm solves for both the object and the probe by minimizing a loss between the model output and measured intensities.
The advantage of iterative ptychography over direct methods is most evident in its ability to recover the object phase with super-resolution~\citep{jiang2018electron}, meaning beyond the diffraction-limited resolution $\lambda/2\alpha$ set by the full convergence angle, and in its ability to recover 3D information about the sample~\citep{maiden2012ptychographic}.
We refer readers to more complete reviews on these topics~\citep{glaeser2013invited, rodenburg2019ptychography, wang2025ptychography, varnavides2026abcs}.

As the value of ptychography has been demonstrated across materials and biological sciences~\citep{Spoth_2017, Yu_2025, Berk_2024}, researchers have placed increasing emphasis on making these methods more broadly applicable. 
This has been in part accomplished through the development of robust open-source packages that make complex computational methods more user-friendly and easily accessible~\citep{Varnavides_2023, Gilgenbach_2025, Lee_2025, quantem}. 
Concurrently, there are efforts to expand to acquisition geometries where iterative ptychography may have previously been intractable. 
For example, \citet{nguyen2024achieving} have demonstrated how ptychography can computationally remove aberrations, leading to a sub-0.5 \AA{} reconstruction in an uncorrected microscope. 

Similarly, there have been demonstrations of super-resolution imaging at low voltage~\citep{allen2023super} including in a scanning electron microscope (SEM)~\citep{humphry2012ptychographic, cao2018image, blackburn2025sub}.
These efforts align with a broader push to expand the capabilities of SEMs to include 4D-STEM and other transmission imaging techniques~\citep{sun2018progress, schweizer2020low, slouf2021powder, parker2022scanning}.
Such efforts are motivated by the many advantages of SEM over STEM.
These include lower cost and accessibility, reduced damage at low accelerating voltage, correlative topographic imaging through secondary electron imaging, and a large specimen chamber for loading large sample arrays and easier \emph{in situ} device studies. 

In this study, we explore the possibility of performing direct and iterative ptychography under nanobeam 4D-STEM imaging conditions.
As in the low accelerating voltage studies described above, our experimental and simulated iterative ptychographic reconstructions return super-resolved images, recovering visible atomic features even at very small convergence angle.
These reconstructions are, however, not uniquely determined by the measured data.
Reconstructing the same dataset twice with different random seeds leads to two reconstructions in which lattice planes are rigidly shifted or the contrast is inverted, which we refer to as inverted polarity. 
Through simulation and theory, we explore the origin of these artifacts.

The source of this ambiguity for crystalline samples below the disk-overlap threshold is that recorded diffraction patterns fix the moduli of each reflection but contain no information about relative phases. 
Real-space overlap between adjacent probe positions does not lift this degeneracy, rendering it a necessary but insufficient criterion for unique convergence.
What can constrain the relative phases is an amorphous substrate, whose diffuse scattering couples reflections that would otherwise never interfere.
However, we find that this phasing is not robust.
The phasing survives only for a forward model that matches the experiment, so realistic model mismatch such as thermal motion, partial source coherence, residual higher-order aberrations, or scan distortion, leads again to degenerate reconstructions.
We demonstrate this explicitly for two such mismatches: constraining the object to be a pure phase object or including thermal motion in the forward model.
Notably, in the simulations the reconstructions are also unstable within a single run, as the recovered lattice position continues to migrate from iteration to iteration.
Taken together, these results indicate that a reconstruction obtained below the disk-overlap threshold is not guaranteed to be unique.

\section{Methods}

\subsection{Experimental data acquisition}
Experimental data in~\cref{fig:dps,fig:ptycho_exp,fig:non_uniform} were collected on the TEAM I microscope at the Molecular Foundry, a modified FEI Titan double-aberration-corrected microscope with a high-brightness Schottky field-emission X-FEG electron source operated in nanoprobe mode.
The data were acquired using a Dectris Arina camera (192 × 192 pixels)~\citep{stroppa2023stem, zambon2023kite}.
The convergence angle was tuned by adjusting the C2/C3 crossover and by switching between apertures. 
We used the 70~µm aperture for 24 and 12~mrad, 40~µm for 6~mrad, 20~µm for 3~mrad, and 10~µm for 1.5~mrad.
The monofocus was modified between scans to keep the probe current at approximately 40~pA, although the spot size remained constant. 
Data were acquired with a 200~µs exposure time.
The probe defocus was adjusted between scans to keep the probe size at approximately 8~\AA{}, and a real-space step size of 0.47~\AA{} was used for all scans.
These collection parameters were chosen to ensure similar probe overlap at all convergence angles and more than sufficient overlap for convergence~\citep{varnavides2026relaxing}.
The total electron fluence was approximately $2\times10^5$~\eA.
Data were acquired in order of decreasing convergence angle.
High-angle annular dark-field (HAADF) images from the same field of view before and after 4D-STEM data acquisition are shown in~\cref{fig:si_haadf}, showing minimal carbon contamination buildup and beam damage from the repeated acquisitions.

\subsection{Experimental data reconstruction}

Datasets for the reconstructions shown in~\cref{fig:dps,fig:ptycho_exp,fig:non_uniform} were binned in reciprocal space by a factor of 2 upon loading.
The initial reciprocal calibration was estimated based on fitting a circle to the bright-field disk.
The rotation calibration was known from previous experiments with the same high tension, detector, and scan rotation, but lower-order aberration coefficients (defocus, two-fold astigmatism magnitude and angle) were estimated using the \texttt{Optuna}~\citep{akiba2019optuna} framework as implemented in the \texttt{DirectPtychography} class of \texttt{quantEM}~\citep{quantem}.
Final parallax reconstructions were upsampled by a factor of 2~\citep{varnavides2026relaxing}.

For the 24~mrad dataset, the reciprocal sampling was further optimized using the iterative ptychographic framework in \texttt{quantEM}~\citep{quantem}. 
This reciprocal sampling was used for iterative ptychographic reconstructions across all other convergence angles.
Final complex-valued object reconstructions were performed with a mixed-state probe model (4 probes).
Ptychographic reconstructions were performed for 50 iterations using the iterative gradient-based reconstruction framework with Adam optimization as implemented in \texttt{quantEM}~\citep{quantem}.

\subsection{Simulations}
\label{sec:methods_sim}

Simulated 4D-STEM datasets were generated with the multislice algorithm~\citep{Cowley_1957, Kirkland_2020} as implemented in \texttt{abTEM}~\citep{madsen2021abtem}, at 300~keV over a $68\times68$~\AA{} field of view sampled on a $340^2$ grid, giving a maximum collection angle of 49~mrad.
The specimen comprises three Au nanoparticles, a [111] single crystal, a [100] single crystal, and a five-fold decahedral twin whose five wedges are each viewed along [110], resting on an 8~\AA{} amorphous carbon film.
The seven crystalline regions so defined ([111], [100], and [110] wedges 1--5) are analyzed independently throughout, and reflect the experiment.
Au nanoparticles on an amorphous carbon film reproduce the specimen of~\cref{fig:dps,fig:ptycho_exp,fig:non_uniform}, and the three orientations present projected lattices of different spacing and symmetry, so that a single dataset provides multiple independent tests of uniqueness.
The decahedral particle additionally tests registry across a boundary.
Its five wedges are twinned on \{111\} planes, so each adjacent pair shares exactly one reflection, whose lattice fringes run continuously across the boundary between them.
The relative position of two neighboring wedges along that shared reflection is therefore fixed by the structure, and a correct reconstruction must reproduce it.
Control datasets omit the carbon film, leaving the particles free-standing.
Scans use a 2~\AA{} step over 1156 probe positions, and three convergence semi-angles of 3, 6 and 12~mrad, with defocus values of 0, 390 and 250~\AA{} respectively, chosen to keep the illuminated area comparable.
The three angles match three of the experimental conditions and straddle the disk-overlap thresholds derived in~\cref{sec:ambiguity}.
Each orientation projects a different set of reflections, so 6~mrad is an informative intermediate.
The [111] particle shows only \{220\} reflections and stays below threshold, while the [100] particle (\{200\}) and the [110] wedges (\{111\}) are above it.
At 12~mrad every region is above threshold and at 3~mrad none are.
Where stated, the forward model was averaged over eight frozen-phonon configurations~\citep{loane1991thermal}.
Poisson noise was applied at fluences between $10^4$ and $10^6$~\eA, alongside a noiseless condition.

Reconstructions use single-slice ePIE~\citep{Maiden_2009} with a known probe, run for 200 iterations.
Note that the best single-slice fit to the multislice exit waves leaves a residual of only 0.25\% of the exit-wave intensity, suggesting a single-slice inverse is sufficient at these thicknesses, even though the forward model is multislice.
We likewise reconstruct a single object state.
A mixed-object model populated its additional modes identically on control data containing no phonons at all, and collapsed the global correlation against ground truth to 0.159.
The extra freedom fits noise rather than thermal motion.

\subsection{Quantifying non-uniqueness}
\label{sec:methods_metrics}

Each simulated condition comprises 4 independent Poisson draws $\times$ 10 initial guesses = 40 reconstructions per convergence angle.
Noiseless conditions are the exception. Only one dataset exists, so they carry a single block of 10 initial guesses, or 50 for the matched on-carbon case.
The ten initial guesses test whether a single measurement uniquely determines the object, while the four draws quantify the variability between nominally similar measurements.

Reconstruction fidelity within each region is measured by the signed peak cross-correlation (``signed\_corr'') of the reconstructed phase against the band-limited ground truth. Its magnitude reports whether the lattice is correct, and its sign whether the contrast is reversed.
The rigid offset is expressed as a ``cell quantile'' $q$, the fraction of the Wigner--Seitz cell lying closer to the origin than the measured offset, so that lattices of different size and shape share a single dimensionless axis.
A value of $q=0$ is perfect registration, and $q=0.5$ is indistinguishable from a random placement.

Two labels distinguish the ways a single run departs from the ground truth.
A region is ``reversed'' where its signed correlation is negative, so that bright columns appear dark.
It is ``ambiguous'' where the polarity is not cleanly defined instead.
This occurs either because the correlation surface carries a competing extremum of opposite sign and comparable magnitude, or because the recovered lattice sits more than a quarter of the Wigner--Seitz cell from its correct position.
The two labels are exclusive, and both require the ground truth to evaluate.
Aggregated over regions, these give the ``reversed'' and ``shifted'' fractions of~\cref{tab:si_summary}, the latter counting regions with $q > 0.25$.

We report ``non-uniqueness'' as the fraction of (draw, region) groups in which the ten initial guesses fail to agree, either differing in polarity, or spreading their offsets over more than a quarter of the Wigner--Seitz cell, that is, $q > 0.25$.
Regions whose lattice was never recovered ($|\mathrm{corr}| < 0.5$) are excluded from this count and reported separately as ``degraded'', so that the number measures genuine ambiguity rather than outright reconstruction failure.
Non-uniqueness therefore compares the ten initial guesses with one another, rather than with the ground truth as the two labels above do.
We emphasize that it can thus be computed without the ground truth, checking whether varying the initial guess changes the reconstructed lattice, and can be performed on experimental data.

\section{Results}

\subsection{Experimental observations}

\begin{figure*}[t]
\centering
\includegraphics[width = \textwidth]{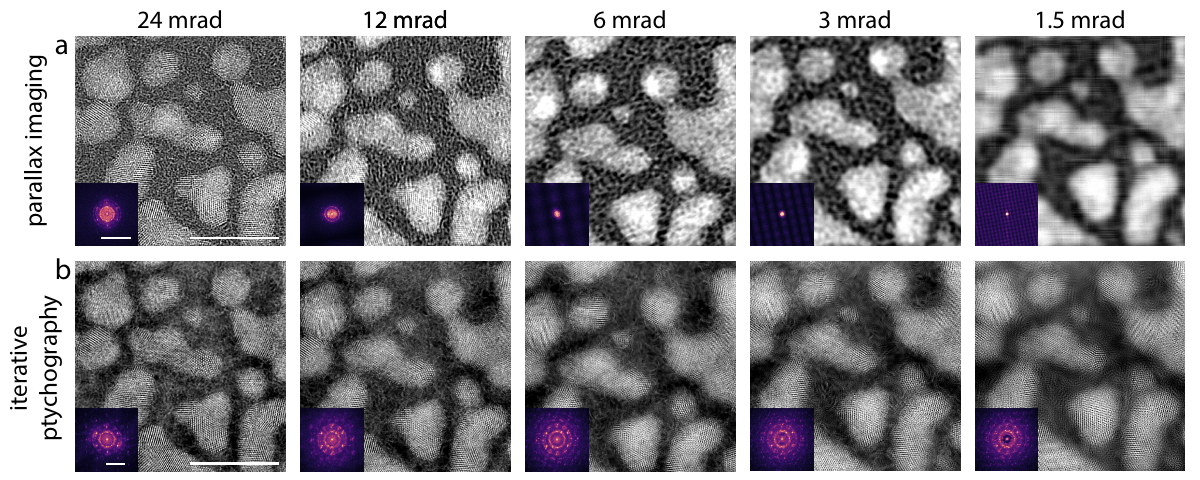}
  \caption{
    (a) Reconstructed object phases obtained using a direct parallax imaging algorithm from 4D-STEM datasets of the same field of view acquired with different convergence angles.
    The information transfer limit decreases with convergence angle. 
    (b) Reconstructed object phases obtained using an iterative ptychography algorithm from 4D-STEM datasets of the same field of view acquired with different convergence angles.
    Although all reconstructions seemingly show atomic-resolution information, only reconstructions with overlap in reciprocal space produce reliable objects.
    Scale bar: 10~nm.
    Inset scale bar: 1/\AA.}
  \label{fig:ptycho_exp}
\end{figure*}

\Cref{fig:dps} shows the mean and individual diffraction patterns from experimental data used in this study.
Data collected with the largest convergence angle (24~mrad) show significant overlap between disks. Individual patterns contain visible lattice structure in the bright-field disk, typical of atomic-resolution defocused ptychography experiments in a corrected microscope~\citep{jiang2018electron, Chen_2021}. 
Disk overlap is significantly reduced at 12~mrad, with features in the bright-field disk being more obscured in part due to the reduced relative sampling of the central beam. 
At 6~mrad, there is no significant overlap in the example diffraction pattern presented. 
The overlap continues to decrease at smaller convergence angles.
As in other 4D-STEM studies~\citep{bustillo20214d}, diffraction patterns at smaller convergence angles improve the signal-to-noise ratio in diffracted disks, making it easier to see reflections at higher scattering angles.

These five scans were reconstructed using parallax imaging as implemented in \texttt{quantEM} (\cref{fig:ptycho_exp}a).
As in conventional imaging, direct phase retrieval transfers information only out to the full convergence angle $2\alpha$, corresponding to a diffraction-limited resolution of $\lambda/2\alpha$~\citep{bennemann2026detective, varnavides2026beyond, Ma_2025}.
This dependence is visible in our reconstructions.
Atomic features fade as the convergence angle is reduced, in both the real-space images and their Fourier transforms, and by 6~mrad all lattice structure has disappeared.
The \{111\} reflection remains nominally inside this limit at 6~mrad, at $2\alpha/\lambda = 0.61$~\AA$^{-1}$ against $g_{111} = 0.42$~\AA$^{-1}$, but transfer so close to the cutoff is too weak to survive the electron dose.
Even at the smallest convergence angles, however, these reconstructions remain useful, accurately locating the nanoparticles and providing the lower-order aberration estimates used to initialize the iterative reconstructions, albeit with an accuracy that degrades as fewer detector pixels sample the bright-field disk.

\Cref{fig:ptycho_exp}b shows the corresponding iterative ptychography reconstructions.
Unlike the direct results, these recover lattice features at every convergence angle, an apparent demonstration of super-resolution.
The reconstructed probes for these datasets are shown in~\cref{fig:si_real_probes,fig:si_reciprocal_probes}.
Spurious artifacts nonetheless appear from 6~mrad downwards, where only the innermost reflections still overlap.
Within the gold islands, contrast reversals leave the bright spots no longer corresponding to atomic columns, while ghost images of the islands appear in the carbon support film, indicating delocalized signal.

\begin{figure*}[t]
\centering
\includegraphics[width = \textwidth]{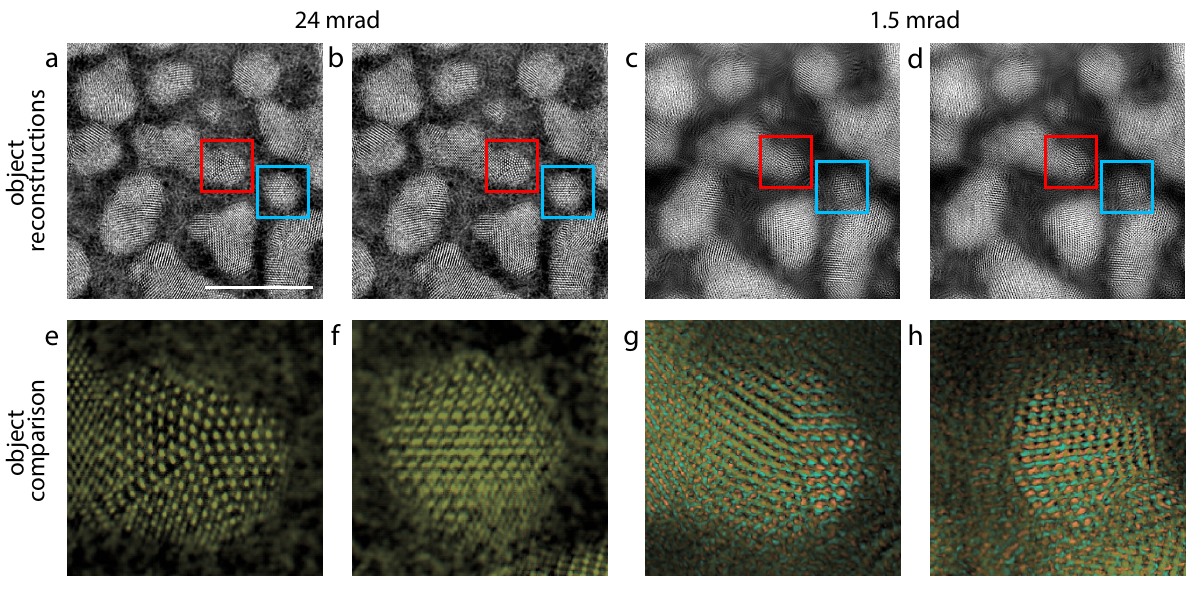}
  \caption{
    Object phase from multiple iterative ptychographic reconstructions of the same dataset using different random seeds controlling the batch ordering of probe positions.
    When the data have overlap in reciprocal space, the reconstruction is well-conditioned to converge to the same object.
    However, without overlap in reciprocal space, the algorithm cannot uniquely converge.
    Scale bar: 10~nm.
  }
  \label{fig:non_uniform}
\end{figure*}

The non-uniqueness of these reconstructions at small convergence angles is highlighted further in~\cref{fig:non_uniform}, which compares pairs of reconstructions of the 24 and 1.5~mrad datasets, run with identical parameters but different random seeds controlling the batch ordering of probe positions.
In the combined images, yellow indicates agreement between the two reconstructions, while pink and teal highlight differences.
At 24~mrad the two are indistinguishable, despite the stochastic nature of the algorithm, but at 1.5~mrad they are not.
The nanoparticle highlighted by the blue box has similar contrast in both reconstructions (\cref{fig:non_uniform}c~\&~d), with atoms appearing bright, but the combined image reveals their atomic positions to be shifted relative to one another, within an unshifted shape envelope.
For the nanoparticle in the red box the contrast is inconsistent across the five twin domains, and it is ambiguous whether atoms are bright or dark, with the difference between the two reconstructions varying from domain to domain.
Such ambiguity makes nanobeam reconstructions unsuitable for studying interfaces and atomic defects in the way that is typical of large convergence angle, atomic-resolution ptychography.

The Fourier transforms of the reconstructed objects in~\cref{fig:ptycho_exp}b naively suggest the reconstruction improves with smaller convergence angles.
The smaller convergence angles have larger signal-to-noise ratio at higher scattering angles, as shown in~\cref{fig:dps}, and potentially improved spatial coherence due to the smaller C2 aperture.
However, the Fourier transform ultimately displays the moduli of the reflections, which a nanobeam measurement can determine well.
By contrast, the relative phases that determine the structure are invisible in this representation.

To confirm that a lack of iterations was not the limitation, the 1.5~mrad dataset of~\cref{fig:ptycho_exp} was additionally reconstructed with the deep generative prior framework, which accelerates convergence~\citep{McCray_2025}.
The resulting object is not significantly different from the pixelated result, and is largely converged after just 10 iterations.
Running for 250 iterations changes neither the locations of the atomic columns nor the object in any significant way (\cref{fig:si_dgp_recon}).
The non-uniqueness we observe is therefore not a matter of insufficient iterations.

Similar behavior is observed at low accelerating voltage.
\Cref{fig:ptycho_60kV} shows iterative multislice ptychographic reconstructions of an evaporated gold on carbon specimen acquired at 60~kV with convergence semi-angles of 30 and 8~mrad.
Since the disk-overlap criterion scales with wavelength, these two angles span the same transition as the 12 and 3~mrad conditions at 300~kV.
At 60~kV ($\lambda = 4.87$~pm) adjacent Au \{111\} disks are separated by 20.7~mrad, so that disks of semi-angle $\alpha$ overlap only for $\alpha > 10.3$~mrad.
The 30~mrad dataset therefore lies well inside the overlapping regime ($2\alpha/\lambda g_{111} = 2.9$), whereas the 8~mrad dataset has no overlap between any Au reflection ($2\alpha/\lambda g_{111} = 0.8$).
As at 300~kV, both reconstructions recover lattice fringes, but only the 30~mrad reconstruction localizes them on the gold particles.
In the 8~mrad reconstruction the fringes extend across the full field of view, including regions of carbon support away from any particle, and the Fourier transform of the object carries a diffuse ring of high-frequency signal alongside the gold reflections.
The reconstructed probes for both datasets are shown in~\cref{fig:si_real_probes_60kV_pair,fig:si_reciprocal_probes_60kV_pair}.

\subsection{Origin of the ambiguity}
\label{sec:ambiguity}

The experimental behavior above has an intuitive origin, which we develop here before turning to simulations.
Consider a crystalline specimen, whose transmission function ($\tilde{t}(\mathbf{k})$) therefore has Fourier support only on the reciprocal lattice,
\begin{equation}
    \tilde{t}(\mathbf{k}) = \sum_{\mathbf{g}} F_{\mathbf{g}}\, \delta(\mathbf{k}-\mathbf{g}),
    \label{eq:crystal}
\end{equation}
where $F_{\mathbf{g}}$ is the complex amplitude of the reflection $\mathbf{g}$.
Illuminating with a probe $\tilde{\psi}(\mathbf{k}) = A(\mathbf{k})\exp[-i\chi(\mathbf{k})]$ positioned at $\mathbf{R}$, where $A(\mathbf{k})$ is the probe-forming aperture of radius $k_0 = \alpha/\lambda$ and $\chi(\mathbf{k})$ the aberration surface, the recorded intensity ($I(\mathbf{R},\mathbf{k})$) is
\begin{equation}
    I(\mathbf{R},\mathbf{k}) = \sum_{\mathbf{g},\mathbf{g}'} F_{\mathbf{g}} F^{*}_{\mathbf{g}'}\,
    \tilde{\psi}(\mathbf{k}-\mathbf{g})\, \tilde{\psi}^{*}(\mathbf{k}-\mathbf{g}')\,
    e^{2\pi i (\mathbf{g}-\mathbf{g}')\cdot\mathbf{R}}.
    \label{eq:intensity}
\end{equation}
Every term carrying phase information about the specimen is a cross term with $\mathbf{g}\neq\mathbf{g}'$, and each such term is supported only where the two shifted apertures intersect~\citep{Rodenburg_1992, Rodenburg_1993}.
Two reflections therefore interfere only when
\begin{equation}
    \alpha > \frac{\lambda\,|\mathbf{g}-\mathbf{g}'|}{2},
    \label{eq:overlap}
\end{equation}
which against the direct beam ($\mathbf{g}'=\mathbf{0}$) reduces to $\alpha > \lambda/2d$ for a lattice spacing $d$.
Note~\cref{eq:overlap} is a purely geometric statement.
It is indifferent to whether the scattering is kinematical or dynamical, and to the accelerating voltage except through $\lambda$.
For Au at 300~keV it places the thresholds at 4.18~mrad for \{111\}, 4.83~mrad for \{200\} and 6.82~mrad for \{220\}.

Below threshold every cross term vanishes and~\cref{eq:intensity} collapses to a diagonal form,
\begin{equation}
    I(\mathbf{R},\mathbf{k}) = \sum_{\mathbf{g}} |F_{\mathbf{g}}|^{2}\,
    \bigl|\tilde{\psi}(\mathbf{k}-\mathbf{g})\bigr|^{2},
    \label{eq:diagonal}
\end{equation}
which depends on neither $\arg F_{\mathbf{g}}$ nor $\mathbf{R}$: the probe-position phase ramp enters each disk but drops out of its intensity, so scanning contributes no phase information either.
The measurement determines the set of reflections $\{\mathbf{g}\}$ and their moduli $\{|F_{\mathbf{g}}|\}$, which is the classical phase problem of crystallography~\citep{Friedel_1913, karle1985}.
Writing $F_{\mathbf{g}} = |F_{\mathbf{g}}|\,e^{i\phi_{\mathbf{g}}}$, any assignment of the phases $\phi_{\mathbf{g}}$ reproduces the data exactly, and two subgroups of that freedom return a physically admissible object:
\begin{align}
    \phi_{\mathbf{g}} &\to \phi_{\mathbf{g}} + 2\pi\,\mathbf{g}\cdot\boldsymbol{\delta}
    &&\text{(rigid translation by } \boldsymbol{\delta}\text{)}, \label{eq:gauge_shift}\\
    \phi_{\mathbf{g}} &\to \phi_{\mathbf{g}} + \pi \quad (\mathbf{g}\neq\mathbf{0})
    &&\text{(contrast reversal)}. \label{eq:gauge_flip}
\end{align}
The first rigidly shifts the lattice by $\boldsymbol{\delta}$.
The second negates the modulated part of the transmission function while leaving its mean untouched, so that atomic columns which were bright become dark.
Note that for a projected lattice with non-collinear lattice vectors, these are genuinely distinct operations.
A translation applies a phase that is linear in $\mathbf{g}$, whereas a reversal applies the same phase $\pi$ to every reflection.
The two coincide only for a single fringe spacing, where a half-period shift and a contrast reversal are indeed equivalent.
Once two non-collinear reflections are present, no single translation can reproduce a contrast reversal.
The reciprocal lattice is closed under addition, so if $\mathbf{g}_1$ and $\mathbf{g}_2$ are each shifted by $\pi$, then $\mathbf{g}_1 + \mathbf{g}_2$ is shifted by $2\pi$ and is left unreversed.
The two are also distinguished in practice, the first by the offset $q$ and the second by the sign of the peak correlation.
These are precisely the two artifacts observed in~\cref{fig:non_uniform}: a lattice of the correct spacing and orientation, sitting in the wrong place or with atomic contrast reversed.

In the notation of~\citet{varnavides2026abcs}, building on~\citet{rose_1977} and~\citet{Hammel_1995}, the aperture-overlap function $\Gamma(\mathbf{q},\mathbf{k}) \equiv \tilde{\psi}^{*}(\mathbf{k})\tilde{\psi}(\mathbf{k}-\mathbf{q}) - \tilde{\psi}(\mathbf{k})\tilde{\psi}^{*}(\mathbf{k}+\mathbf{q})$ vanishes identically for $|\mathbf{q}| > 2k_0$, so direct methods such as SSB, OBF and parallax have zero transfer at every reflection beyond twice the aperture.
This can be seen in~\cref{fig:ptycho_exp}a, where lattice contrast disappears from the parallax reconstructions by 6~mrad.
Iterative algorithms are not bound by that limit, because they exploit the nonlinear terms in the forward model, and this is the origin of their super-resolution~\citep{jiang2018electron, Chen_2021}.
What~\cref{eq:diagonal} shows is that below threshold those nonlinear terms retain the moduli but carry no relative phase.
Iterative ptychography therefore continues to place signal at the correct spatial frequencies at arbitrarily small convergence angle, while the information needed to correctly position that lattice is absent from the data.
This is why the reconstructions in~\cref{fig:ptycho_exp}b appear to show a lattice, and why their Fourier transforms appear to improve with decreasing convergence angles.

The degeneracy of~\cref{eq:gauge_shift,eq:gauge_flip} is exact only for an isolated crystal.
A non-periodic substrate breaks it.
Since the total transmission function in real space is a product of the specimen $V_{\mathrm{Au}}(\mathbf{r})$ and substrate $V_{\mathrm{C}}(\mathbf{r})$ terms,
\begin{equation}
    T(\mathbf{r}) = e^{i\sigma V_{\mathrm{Au}}(\mathbf{r})}\, e^{i\sigma V_{\mathrm{C}}(\mathbf{r})},
    \label{eq:product}
\end{equation}
in reciprocal space it is a convolution, $\tilde{T} = \tilde{T}_{\mathrm{Au}} \ast \tilde{T}_{\mathrm{C}}$.
Amorphous carbon scatters diffusely, with continuous support and no Bragg peaks, so the convolution gives the Au reflections amplitude in the gaps between the disks.
Writing $\tilde{T}_{\mathrm{C}}(\mathbf{k}) = \delta(\mathbf{k}) + \tilde{c}(\mathbf{k})$, where $\delta(\mathbf{k})$ is the unscattered component and $\tilde{c}(\mathbf{k})$ the diffuse amplitude scattered by the carbon, every reflection acquires amplitude $F_{\mathbf{g}}\tilde{c}$ outside its own disk.
For a detector point inside disk $\mathbf{g}$ but outside disk $\mathbf{g}'$, the surviving cross term is proportional to
\begin{equation}
    F_{\mathbf{g}} F^{*}_{\mathbf{g}'}\, \tilde{\psi}(\mathbf{k}-\mathbf{g})\,
    \tilde{c}^{*}(\mathbf{k}-\mathbf{g}')\,
    e^{2\pi i (\mathbf{g}-\mathbf{g}')\cdot\mathbf{R}},
    \label{eq:substrate_cross}
\end{equation}
which depends on the relative phase and on probe position.
The substrate carries phase information between reflections that would otherwise never interfere.
It enters~\cref{eq:substrate_cross} only through $\tilde{c}$, so this phasing is first order in the substrate scattering, whereas above threshold the reflections interfere directly and the phasing is zeroth order in it.

Finally, we note what real-space overlap contributes, since this criterion is most often quoted.
For a probe of diameter $d_{\mathrm{p}}$ and scan step $s$, the number of probe positions illuminating a given point is
\begin{equation}
    R = \frac{A_{\mathrm{probe}}}{s^{2}} = \frac{\pi/4}{(1-o)^{2}},
    \label{eq:redundancy}
\end{equation}
where $o = 1 - s/d_{\mathrm{p}}$ is the linear overlap, so that $R$ is set by the overlap alone and not by the size of the probe.
Our experiments, with an 8~\AA{} probe and a 0.47~\AA{} step, give $o \approx 0.94$ and $R \approx 228$.
Such redundancy constrains the object wherever the illumination varies from one position to the next.
Below threshold~\cref{eq:diagonal} carries no dependence on $\mathbf{R}$ at all, so there is no position-dependent information for redundancy to leverage, however large $R$ is made.

\subsection{Phase-free negative control}
\label{sec:stitching}

\begin{figure*}[t]
\centering
\includegraphics[width = \textwidth]{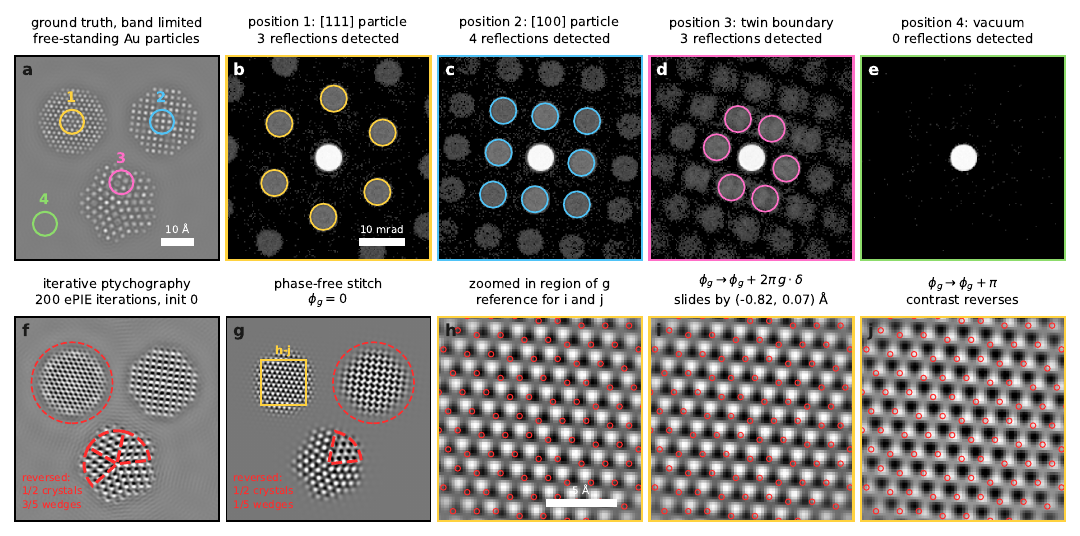}
  \caption{
    Negative control reproducing the two failure modes from a phase-free reconstruction algorithm.
    Simulated free-standing Au at 3~mrad and $10^5$~\eA, so that no pair of Bragg disks overlaps.
    (a) Band-limited ground truth with four probe positions marked.
    (b--e) The diffraction patterns recorded at each marked probe position, with detected reflections circled.
    That list of positions and intensities constitutes the entire measurement passed to the algorithm.
    (f) ePIE reconstruction of the same specimen for comparison.
    (g) A real-space image synthesized directly from the detected positions using~\cref{eq:stitch}, assigning every reflection a zero phase.
    Both (f) and (g) invert regions, though not the same ones, since which regions flip is arbitrary in both. Inverted regions are marked in red.
    (h) The stitched image of (g), reproduced as the reference for (i) and (j). (i, j) The two operations of~\cref{eq:gauge_shift,eq:gauge_flip} applied to it: a rigid translation slides the lattice off the atom columns by 0.82~\AA, and a global $\pi$ shift reverses contrast. The ground truth locations of atomic columns are marked with red circles.
  }
  \label{fig:stitching}
\end{figure*}

If the argument of~\cref{sec:ambiguity} is correct, then the artifacts we observe are not a property of ePIE, or of iterative optimization, or of any particular regularization.
Instead, they follow from the measurement itself.
We test this by constructing a reconstruction algorithm that explicitly performs no phase retrieval.

For each probe position $\mathbf{R}$ the diffraction pattern is reduced to a list of detected reflections and their intensities.
Each reflection is assigned a phase $\phi_{\mathbf{g}}$, which the measurement does not constrain and which we are therefore free to set.
We accumulate a real-space image ($C(\mathbf{r})$) using
\begin{equation}
    C(\mathbf{r}) = \sum_{\mathbf{R}} w(\mathbf{r}-\mathbf{R})
    \sum_{\mathbf{g}\in G(\mathbf{R})} \sqrt{I_{\mathbf{g}}(\mathbf{R})}\,
    \cos\bigl(2\pi\,\mathbf{g}\cdot\mathbf{r} + \phi_{\mathbf{g}}\bigr),
    \label{eq:stitch}
\end{equation}
and display it as $C(\mathbf{r})/W(\mathbf{r})$ with $W(\mathbf{r}) = \sum_{\mathbf{R}} w(\mathbf{r}-\mathbf{R})$, where $w$ is the probe intensity envelope and $G(\mathbf{R})$ the set of reflections detected at $\mathbf{R}$.
Reflections are clustered into families across the scan, so that a single phase applies to each family, in the manner of orientation and flowline mapping of nanobeam data~\citep{panova2019diffraction, ophus2022automated}.
Assigning phases independently at every position instead produces visible incoherence, confirming that a phase must at least be consistent from one position to the next for a lattice to appear at all.
The reference map takes $\phi_{\mathbf{g}} = 0$ throughout.

\Cref{fig:stitching} shows the result at 3~mrad on free-standing Au nanoparticles.
Both the ePIE reconstruction and the phase-free map recover the lattice, and both return regions with reversed contrast, but notably not the same regions.
ePIE reverses [111] and three of the five twin wedges, at a global correlation of 0.622, while the stitched map reverses [100] and one wedge, at 0.517.
Neither is more correct than the other, because the unmodified map with $\phi_{\mathbf{g}}=0$ is as good a solution as any other choice.
Panels (i) and (j) apply~\cref{eq:gauge_shift,eq:gauge_flip} directly to the reference map of panel (h): a translation of $\boldsymbol{\delta} = (-0.82, 0.07)$~\AA{} slides the lattice off the atom columns, and a global $\pi$ shift reverses contrast, with every recorded intensity unchanged in both cases.

An algorithm containing no phase retrieval thus fails in exactly the same two ways as ePIE, and to a similar extent.
We take this as evidence that the failure belongs to the measurement rather than to the algorithm, and that below threshold no phase retrieval is in fact taking place.

\subsection{Sensitivity of the substrate phasing channel}
\label{sec:uniqueness}

\begin{figure*}[t]
\centering
\includegraphics[width = \textwidth]{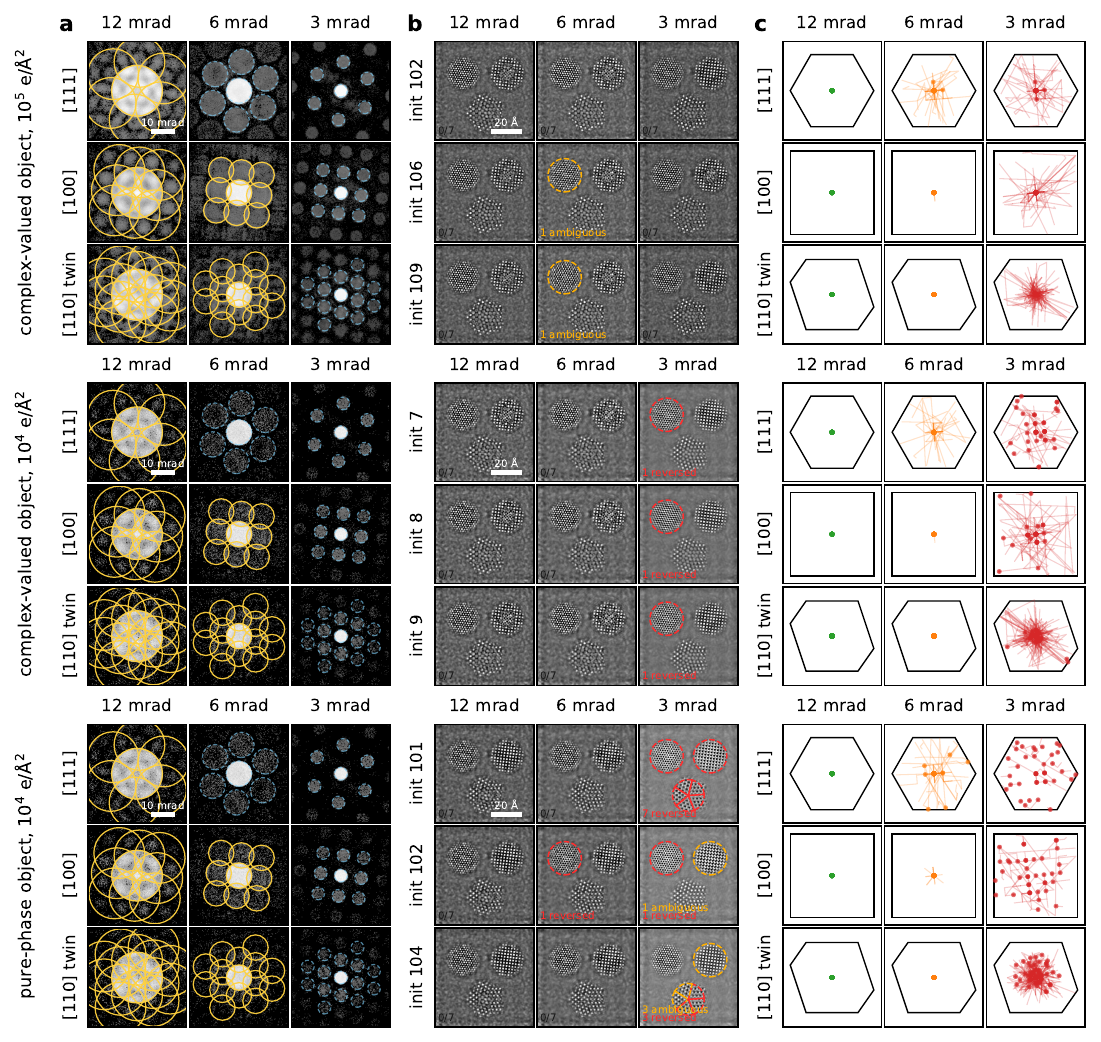}
  \caption{
    Below the disk-overlap threshold, a matched forward model at sufficient dose is unique.
    All rows use a multislice forward model with the particles on amorphous carbon, and each panel represents 40 reconstructions (4 Poisson draws $\times$ 10 initial guesses).
    Top: a complex-valued object at $10^5$~\eA{} recovers all seven regions unanimously at every angle, with non-uniqueness 0.00 even at 3~mrad where no disks overlap.
    Middle: dropping the dose to $10^4$~\eA{} breaks this at 3~mrad only (0.21), while 6 and 12~mrad remain unanimous.
    Bottom: holding the dose at $10^4$~\eA{} and constraining the object to pure phase raises 3~mrad to 0.82 and puts visible ambiguity into the 6~mrad column (0.11).
    Panels are (a) diffraction data at the row's dose, with disks outlined solid where two overlap and dashed where they do not;
    (b) three initial guesses of one dataset, chosen as the most visibly failed, with regions outlined red where the contrast is reversed and amber where the polarity is ambiguous, both as defined in~\cref{sec:methods_metrics};
    (c) per-iteration offset trails for every run inside the Wigner--Seitz cell, which collapse to the origin where the reconstruction is unique and fill the cell where it is not.
    The outlines in (b) describe a single run against the ground truth, whereas the non-uniqueness quoted for each row describes the ten initial guesses against one another.
    The polarity of every individual run is shown separately in~\cref{fig:si_sparklines}.
  }
  \label{fig:sweep}
\end{figure*}

\Cref{sec:ambiguity} predicts that below threshold an isolated crystal is ambiguous, and that an amorphous substrate can remove the ambiguity.
We test both predictions against the simulations described in~\cref{sec:methods_sim}.
\Cref{fig:si_substrate} tests the substrate importance directly.
With the particles resting on amorphous carbon, and a forward model matched to the reconstruction, the ten initial guesses agree with one another on every one of the four 3~mrad datasets at $10^5$~\eA, with a mean $|\mathrm{corr}|$ of 0.897 across all 40 reconstructions.
Removing the carbon substrate makes all of the reconstructions ambiguous.
Every group of initial guesses in the free-standing row disagrees, and 44\% of regions come back with reversed contrast.
This holds at every dose we tested, including the noiseless case.
Infinite dose therefore does not help, which marks the free-standing failure as a structural limit rather than a counting-statistics one, exactly as~\cref{eq:diagonal} requires.
The mechanism is visible in the diffraction data itself.
Free-standing patterns show isolated disks on an empty background, while supported patterns show diffuse scattering filling the gaps between them.

Nanobeam iterative ptychography below the overlap threshold is therefore possible in principle.
However, \cref{fig:sweep} shows how sensitive it is to experimental artifacts such as Poisson noise and model mismatch.
Each row changes one variable, with the particles on a carbon substrate throughout.
Dropping the dose from $10^5$ to $10^4$~\eA{} breaks uniqueness at 3~mrad, while 6 and 12~mrad remain unanimous.
Additionally, constraining the object to be a pure phase object, a common and physically motivated regularization~\citep{Vulovic_2014}, makes 3~mrad roughly four times worse and introduces visible ambiguity in the [111] particle at 6~mrad.
Averaging the forward model over eight frozen-phonon configurations~\citep{loane1991thermal} raises the ambiguity and degrades the reconstruction at the same time, with mean $|\mathrm{corr}|$ falling from 0.793 to 0.696 and the degraded fraction rising from 0.05 to 0.16.
Thermal motion also sets the achievable resolution in ptychography more generally~\citep{Chen_2021}.
\Cref{fig:si_phonons} shows this case, and~\cref{tab:si_summary} collects every condition tested.

One feature of~\cref{tab:si_summary} is worth highlighting.
Free-standing particles at 6~mrad are close to unanimous, at 0.07 and 0.04 for the two lowest doses, rather than the 1.00 seen at 3~mrad.
Ambiguity therefore largely resolves once the \{111\} and \{200\} reflections begin to overlap, even with no substrate present, which is what~\cref{eq:overlap} requires.
The effect tracks disk overlap rather than the substrate, and the substrate matters only where overlap is absent.
We note also that the pure phase object condition is not monotonic in dose, at 0.71, 0.21, 0.25 and 0.82 from noiseless down to $10^4$~\eA.

The two gauge freedoms are also not observed equally in these simulations.
Across the 8925 region measurements behind~\cref{tab:si_summary}, 488 return with reversed contrast and 223 sit more than a quarter of the Wigner--Seitz cell from their correct position, so polarity errors outnumber translational ones by about two to one.
Our twinned particle carries a coupling channel of its own, since adjacent wedges share their twin-plane reflection and their contributions therefore interfere in a common disk at any convergence angle.
While this is a genuine phasing channel, it is a weak one.
It constrains only a single relative phase, and it is second order in the Bragg amplitudes, whereas the interference supplied by disk overlap is first order in the direct beam, which carries 83\% of the pattern intensity at 3~mrad.
Small registry errors across our twin boundaries are therefore not evidence that a below-threshold reconstruction places lattices correctly.
The experimental twin domains in~\cref{fig:non_uniform} do vary in contrast from one to the next, which we attribute to model mismatch masking a channel this weak, just as it masks the substrate channel.

\Cref{fig:si_sparklines} resolves these reconstructions by initial guess, and two features of it matter.
The first is that mean $|\mathrm{corr}|$ stays between 0.79 and 0.82 in all three conditions, suggesting the lattice is always recovered, with the correct spacing and orientation.
Only its sign and origin scatter, so no single reconstruction looks wrong.
This is why we report agreement between repeated reconstructions rather than a quality metric.
The second is the spread between nominally identical measurements.
At 3~mrad and $10^4$~\eA{} with a pure phase object, the four Poisson draws disagree at rates of 1.00, 1.00, 1.00 and 0.29.
Three datasets are non-unique in every region, while the fourth is mostly well behaved, even though all four were acquired under identical conditions.
This is directly visible in~\cref{fig:si_sparklines}c, where the fourth block of rows is markedly cleaner than the first three.

The offset trails in~\cref{fig:sweep}c show the same effect during a single reconstruction.
Where the answer is ambiguous, the recovered lattice position never settles, but migrates from one iteration to the next.
In the unanimous conditions the trails collapse to the origin instead.
This follows from the structure of the degeneracy.
The residual freedom of~\cref{eq:gauge_shift,eq:gauge_flip} lies along a direction the data do not constrain, so moving along it costs nothing in the loss.
An optimizer can therefore reach a stationary loss without the specimen having been determined, and whether the solution then drifts or sits still is governed by the regularization rather than by the measured data.

\section{Discussion}

\subsection{Reconciling the simulations with the experiment}

The simulations of~\cref{sec:uniqueness} find that below-threshold reconstructions on an amorphous substrate are unique at electron doses $10^5$~\eA{} and above, whereas our experimental reconstructions from data acquired at a comparable dose are ambiguous below 6~mrad.
Neither electron dose nor real-space sampling can account for the discrepancy, considering the simulations are an order of magnitude less redundant than the experimental acquisition.
We conclude that what fails in the experiment is the match between the forward model and the measurement under typical experimental conditions.

The simulations isolate the individual contributions to that mismatch, and each is on its own sufficient.
Constraining the object to be a pure phase object raises 3~mrad non-uniqueness to 0.82 at fixed dose, while averaging over frozen phonons raises it to 0.32 while also degrading quality.
An experiment carries multiple model mismatches at once.
Our reconstructions refine four mixed probe states~\citep{Thibault_2013, Chen_2020}, which introduces an object--probe ambiguity absent from the simulations~\citep{Odstrcil_2016}.
Residual higher-order aberrations, scan distortion, partial coherence and detector response~\citep{Levin_2021} all contribute further.
It is therefore evident that the dose thresholds presented are guidelines for marginal stability in a best-case scenario only, and that realistic model error means that those values do not apply to experimental data.
The uniqueness demonstrated in the top row of~\cref{fig:sweep} should accordingly be read as an upper bound on what is achievable, not as a specification that an experiment can be designed to meet.

The frozen-phonon case deserves particular emphasis, because it is the hardest to diagnose.
Unlike every other mismatch tested, thermal motion degrades reconstruction quality as well as introducing ambiguity.
The resulting graininess reads naturally as insufficient dose, and invites a longer exposure or heavier regularization as the remedy.
It gives no indication that the lattice is also sitting in the wrong place.

\subsection{Practical considerations}

The small convergence angle reconstructions in~\cref{fig:ptycho_exp}b do carry spatial frequencies well beyond the limit of the direct reconstructions in~\cref{fig:ptycho_exp}a.
However, they do not carry the necessary phase information to position these spatial frequencies uniquely.
They are therefore poorly suited to high-resolution characterization questions, such as the identity of a point defect or the registry across an interface.

What nanobeam data do support well is strain and orientation mapping.
These datasets in particular are well suited to studying orientation relations, where automated crystal orientation mapping reaches the same information by a more directly interpretable route~\citep{rauch2010automated, cautaerts2022free, ophus2022automated}.
Flowline mapping likewise uses nanobeam diffraction to map crystalline domains in semi-crystalline polymers~\citep{panova2019diffraction}.
Our nanobeam reconstructions are close relatives of these methods, and the stitching construction of~\cref{sec:stitching} makes the relationship explicit.
These methods report the orientation and spacing of crystalline planes across a field of view without establishing where each plane sits.
The difficulty arises only when such a reconstruction is read as an atomic-resolution image.

The challenge of non-unique phasing in nanobeam ptychography is analogous to long-standing phase and transfer problems in electron microscopy.
In high-resolution TEM, contrast reversals, zero crossings, and image delocalization limit direct quantitative image interpretation, motivating the development of focal-series and exit-wave reconstruction methods~\citep{coene1992image}. 
Early iterative phase-retrieval algorithms, including the Gerchberg--Saxton algorithm, established a framework for recovering phase from intensity measurements, but such methods often suffer from convergence to local minima and non-unique solutions~\citep{gerchberg1972practical, latychevskaia2012holography, shechtman2015phase}. 
Related projection algorithms, including hybrid input--output, became central to coherent diffraction imaging by incorporating constraints to improve convergence~\citep{fienup1978reconstruction, Fienup_1982}. 
Nevertheless, practical limitations can still produce ambiguities in CDI reconstructions, complicating quantitative structure determination in a manner analogous to the limitations observed here~\citep{de1996direct, ophus2012guidelines, latychevskaia2012holography, shechtman2015phase, Wu2005, Wu2006}.

Note that none of this is an argument against nanobeam data acquisition.
At lower accelerating voltages the optics themselves limit the usable angle, through both spatial coherence and residual aberrations, so nanobeam conditions are often not a matter of choice.
Where the angle can be chosen, overlap is in any case not the only consideration, since a larger convergence angle also gives better depth sensitivity for multislice ptychography, which can be extended further with aperture synthesis~\citep{gao2024central, allars2025depth, dong2025sub}.
The narrower point is that where the diffracted disks do not overlap, an iterative reconstruction should not be read as an atomic-resolution image without independent evidence that it is unique.

\section{Summary and outlook}

In this work we highlight, through experiment and simulation, the challenge in performing ptychographic reconstructions at low convergence angles.
When the convergence angle is small enough that no diffracted disks overlap, the recorded intensities do not carry relative phase information.
In this case, reconstructed ptychographic objects can show rigid translations of the lattice and reversals of contrast. 
These ambiguities are reproduced by a stitching procedure that performs no phase retrieval at all.
Real-space overlap between adjacent probe positions does not constrain either.

An amorphous substrate gives the Bragg beams support in the gaps between disks and so couples reflections that would otherwise never interfere, and with such a substrate our simulated reconstructions below the overlap threshold are indeed unique.
The practical difficulty is that this phasing channel is fragile.
It requires sufficient dose, and it requires a forward model that matches the experiment.
At fixed dose, either a pure phase object constraint or thermal motion in the forward model is on its own enough to destroy it.
Our own experiments, at a dose and redundancy well beyond the simulated thresholds, are ambiguous.
We attribute this to model error, for which an experiment contains several sources simultaneously.

Two consequences follow for practice.
First, because the lattice is recovered correctly throughout and only its position and polarity scatter, these failures are invisible in any single reconstruction.
The diagnostic is to vary the initial guess or batch ordering of probe positions and see whether the reconstruction is robust.
Second, for experiments at low accelerating voltage or without a corrector, other 4D-STEM approaches may be more appropriate with current reconstruction algorithms.
It may yet be possible to resolve the ambiguity with additional constraints based on the sample geometry, or by deliberately introducing a diffusely scattering support.
A general iterative ptychographic framework, while extremely valuable when properly phased in reciprocal space, cannot be relied upon under nanobeam conditions.

\section{Declaration of competing interests}

The authors declare no competing interests.

\section{Declaration of generative AI and AI-assisted technologies use}

During the preparation of this work the authors used Claude Opus 5 in order to write and debug scripts for running numerical simulations, generate code for data visualization and figure plotting, and proofread the manuscript text to improve language structure and clarity.
After using this tool, the authors reviewed the validity and reproducibility of the generated figures and take full responsibility for the content of the published article.

\section{Data statement}
The experimental datasets supporting the findings of this study are publicly available on \href{https://doi.org/10.5281/zenodo.22132838}{\texttt{Zenodo}}.
All simulation, reconstruction, and analysis notebooks used to generate the figures are available on \href{https://github.com/smribet/nanobeam_ptycho}{\texttt{GitHub}}.

\section{Author contributions}
S.M.R.\ and M.D.\ collected the experimental datasets.
S.M.R.\,  M.D.\, and F.A.\ performed the experimental reconstructions.
A.R.C.M.\ performed the deep generative prior experimental reconstructions.
W.P.M.d.K.\ and G.V.\ designed and performed the simulations, implemented the phase-free stitching control, and derived the analytic results.
C.S.A.\ and C.O.\ contributed to the conception and design of the study.
The first draft of the manuscript was written by S.M.R., M.D.\ and G.V., and all authors commented on previous versions.
All authors read and approved the final manuscript.

\section{Acknowledgments}
The authors thank Jim Ciston and David A. Muller for the helpful discussions.

\section{Funding sources}
Work at the Molecular Foundry was supported by the Office of Science, Office of Basic Energy Sciences, of the U.S. Department of Energy under Contract No. DE-AC02-05CH11231, under user proposal numbers MFP-09619, MFP-09942.
We thank Diamond Light Source for access and support in use of the electron Physical Science Imaging Centre (Instrument E02 and proposal number MG45949) that contributed to the results presented here.
Work at TU Delft is part of the research program Foundations for electron-beam metrology and inspection with file number KICH2.V4C.22.001 which is financed by ASML and the Dutch Research Council (NWO).
A.R.C.M. and C.O. thank the Accelerated Materials Design \& Discovery Program in the Toyota Research Institute for funding support.

\bibliographystyle{elsarticle-harv} 
\bibliography{reference}

\newpage
\clearpage

\setcounter{section}{0}
\renewcommand\thesection{S\arabic{section}.}

\setcounter{page}{1}
\renewcommand\thepage{S.\arabic{page}}

\setcounter{figure}{0}
\renewcommand\thefigure{S.\arabic{figure}}

\setcounter{equation}{0}
\renewcommand{\theequation}{S\arabic{equation}}

\providecommand{\theHsection}{}\renewcommand{\theHsection}{S\arabic{section}}
\providecommand{\theHpage}{}\renewcommand{\theHpage}{S\arabic{page}}
\providecommand{\theHfigure}{}\renewcommand{\theHfigure}{S\arabic{figure}}
\providecommand{\theHtable}{}\renewcommand{\theHtable}{S\arabic{table}}
\providecommand{\theHequation}{}\renewcommand{\theHequation}{S\arabic{equation}}

\onecolumn
\begin{center}
\textbf{{\Large Real-space overlap is not enough: ambiguity in nanobeam iterative ptychography}}

\end{center}

{\large Stephanie M. Ribet$^{1,\dagger}$, Mohsen Danaie$^{2,\dagger}$, Willem P.M. de Kleijne$^{3,\dagger}$, Arthur R. C. McCray$^{4}$, Frederick Allars$^{2}$, Christopher S. Allen$^{2}$, Colin Ophus$^{4}$, Georgios Varnavides$^{3}$}

~\hspace{-2em} $^{1}$ National Center for Electron Microscopy, Molecular Foundry, Lawrence Berkeley National Laboratory, Berkeley, 94720, CA, United States of America

~\hspace{-2em} $^{2}$ Electron Physical Science Imaging Centre, Diamond Light Source, Harwell Science and Innovation Campus, Didcot, OX11 0DE, Oxfordshire, United Kingdom

~\hspace{-2em} $^{3}$ Department of Imaging Physics, Delft University of Technology, Lorentzweg 1, Delft, 2628 CJ, The Netherlands

~\hspace{-2em} $^{4}$ Department of Materials Science and Engineering, Stanford University, Stanford, 94305, CA, United States of America

~\hspace{-2em} $^{\dagger}$These authors contributed equally to this work

~\hspace{-2em} Email addresses: sribet@lbl.gov, mohsen.danaie@diamond.ac.uk, g.varnavides@tudelft.nl

\begin{figure*}[ht]
\centering
\includegraphics[width = \textwidth]{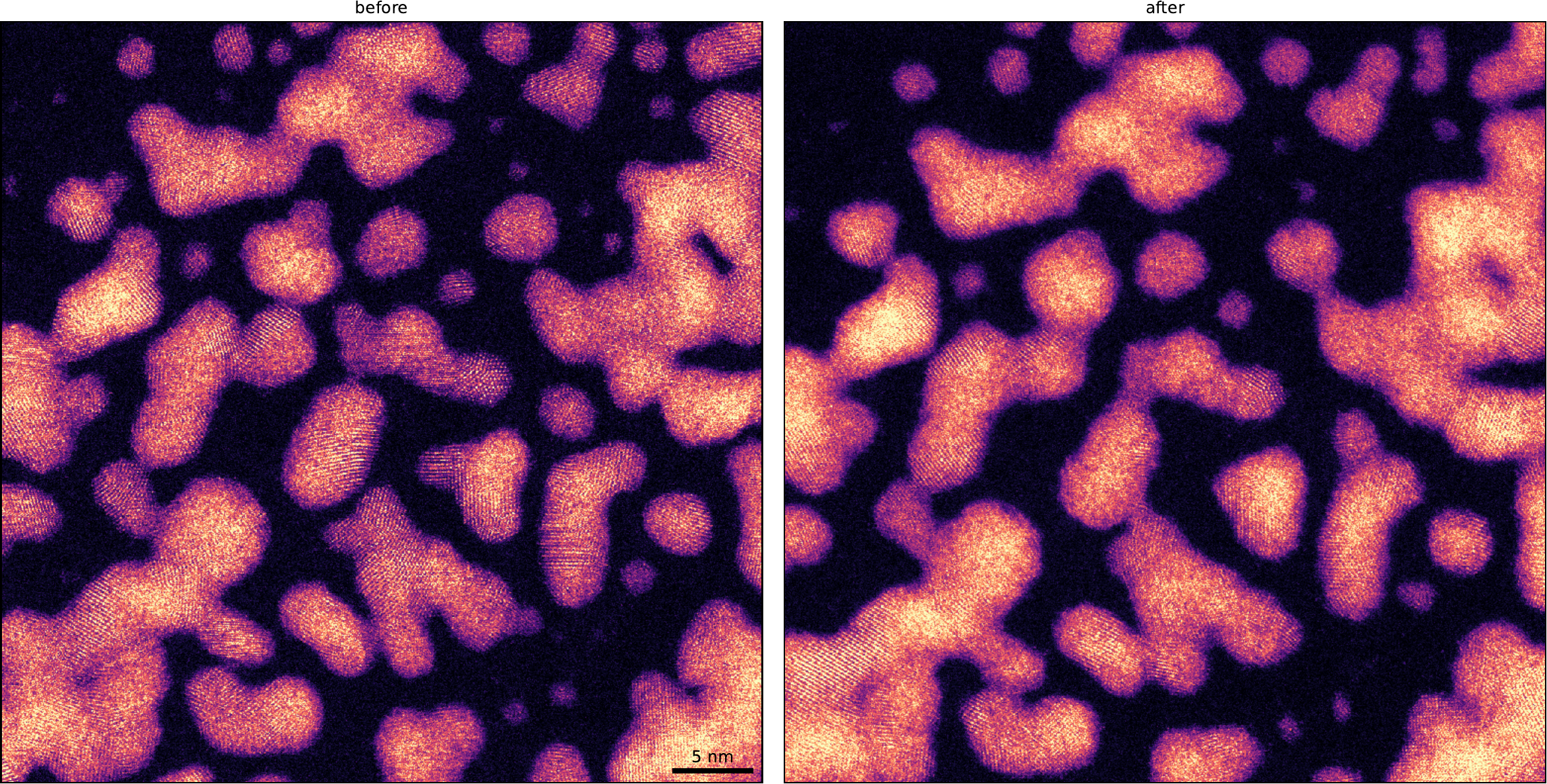}
  \caption{
    High-angle annular dark-field (HAADF) images taken before and after 4D-STEM data acquisition for data in~\cref{fig:dps,fig:ptycho_exp,fig:non_uniform}. 
}
  \label{fig:si_haadf}
\end{figure*}

\begin{figure*}[ht]
\centering
\includegraphics[width = \textwidth]{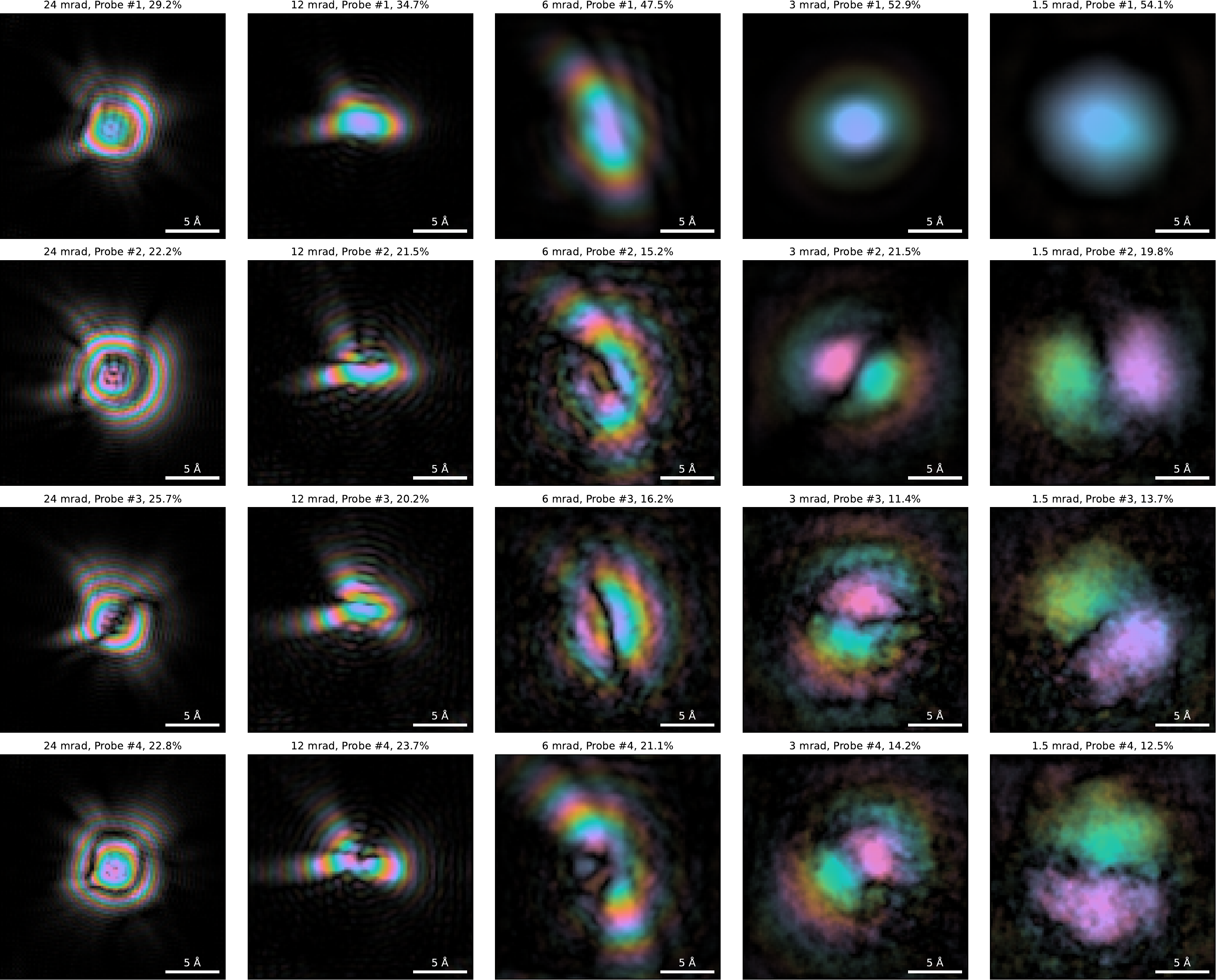}
  \caption{
    Real-space reconstructed probes from reconstructions in~\cref{fig:ptycho_exp}b with associated mode weights.
    }
  \label{fig:si_real_probes}
\end{figure*}

\begin{figure*}[ht]
\centering
\includegraphics[width = \textwidth]{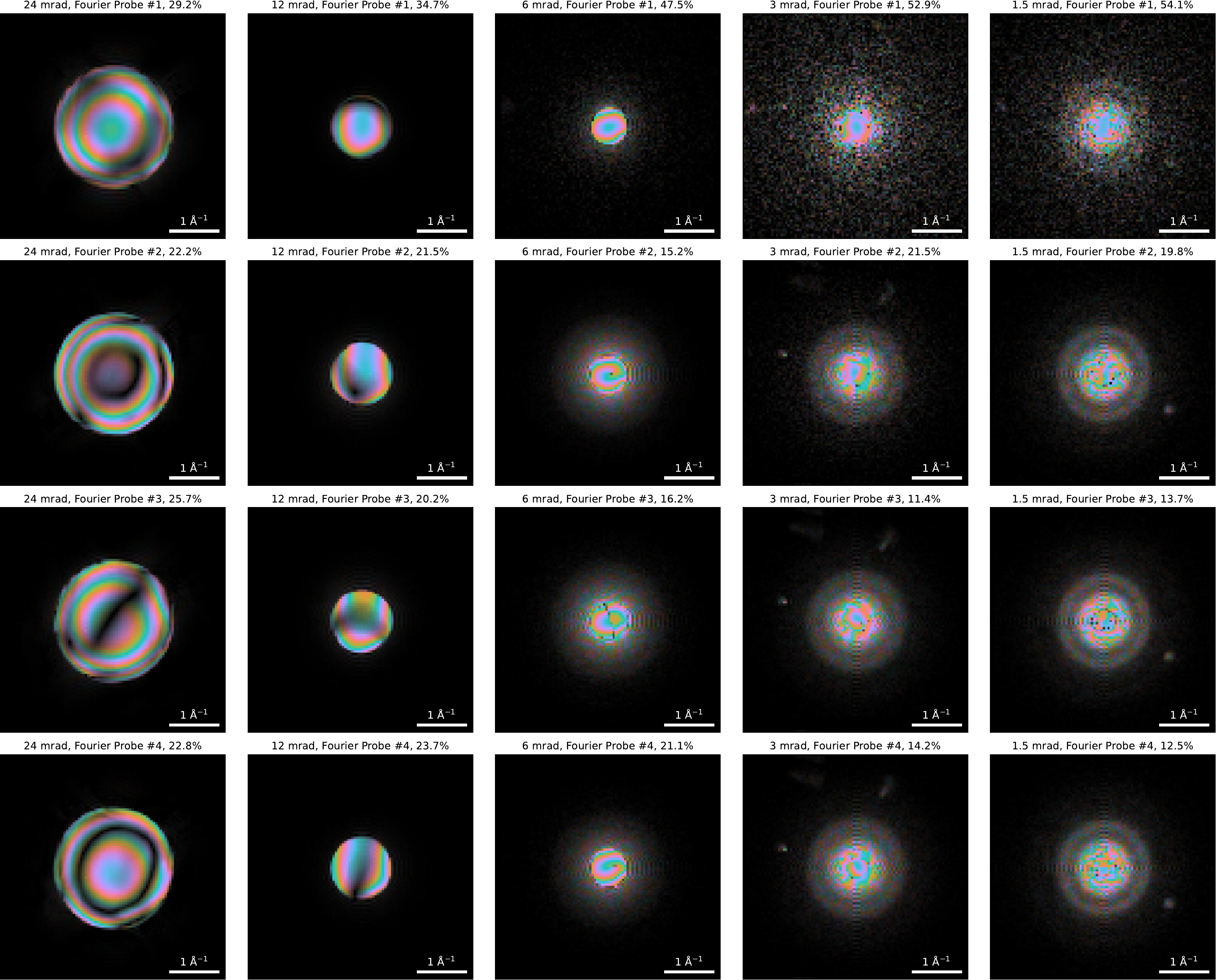}
  \caption{
    Reciprocal-space reconstructed probes from reconstructions in~\cref{fig:ptycho_exp}b with associated mode weights.
    }
  \label{fig:si_reciprocal_probes}
\end{figure*}

\begin{figure*}[ht]
\centering
\includegraphics[width = \textwidth]{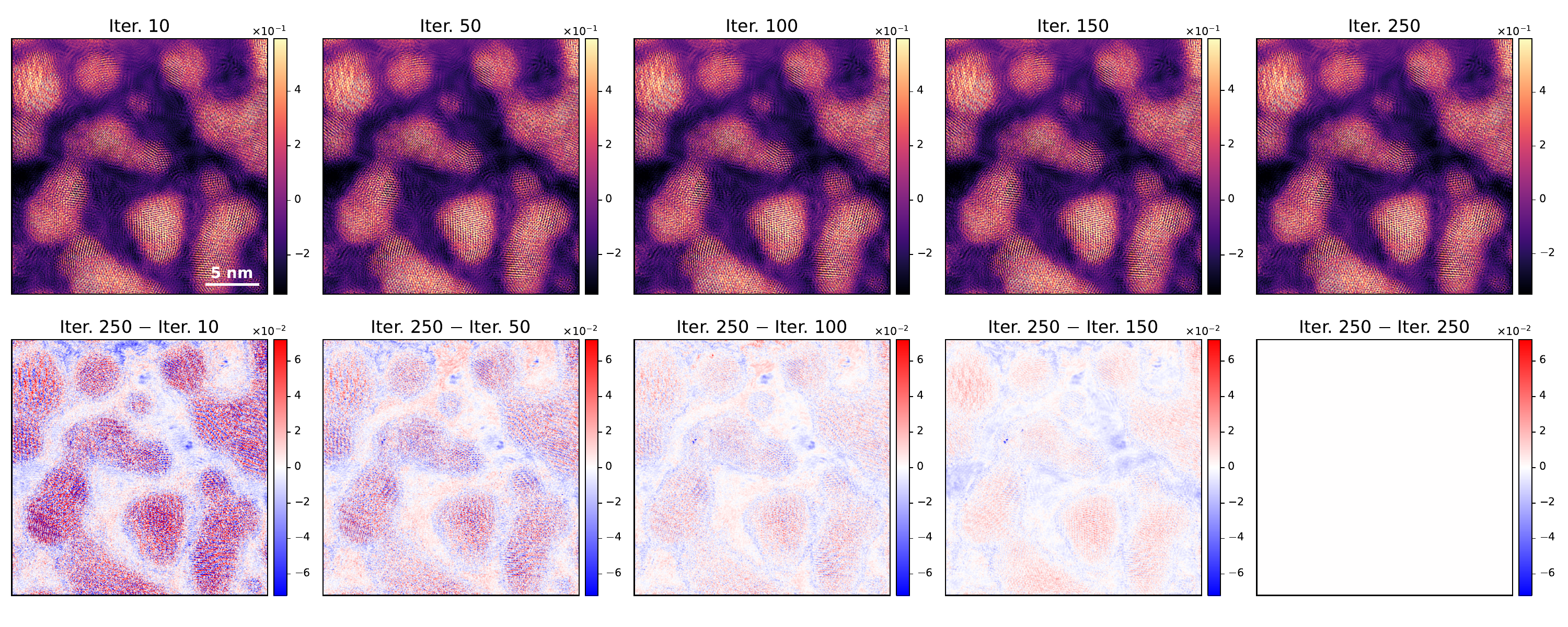}
  \caption{
    Deep generative prior reconstruction of the 1.5~mrad dataset shown in~\cref{fig:ptycho_exp}.
    The reconstruction is effectively fully converged after 10 iterations, and further iterations alter neither the locations of atomic columns nor the overall quality of the reconstructed object.
    }
  \label{fig:si_dgp_recon}
\end{figure*}

\begin{figure*}[ht]
\centering
\includegraphics[width = \textwidth]{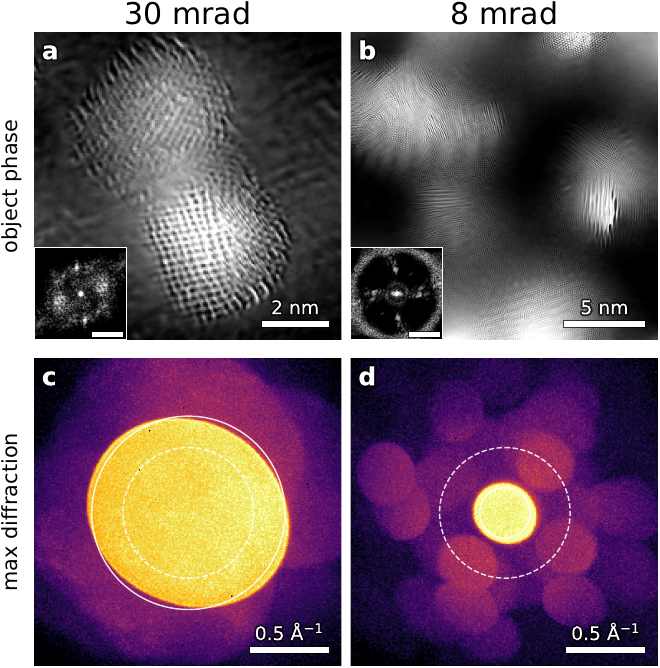}
  \caption{
    (a, b) Reconstructed object phases obtained using an iterative multislice ptychography framework from 60~kV 4D-STEM datasets acquired with convergence semi-angles of (a) 30~mrad and (b) 8~mrad; insets show the Fourier transform of each reconstructed phase.
    (c, d) Maximum diffraction pattern over the corresponding scan.
    Solid circles mark the bright-field disk edge ($\alpha/\lambda$) and dashed circles the Au \{111\} disk spacing ($g_{111}$).
    Diffracted disks interfere with the direct beam only where the dashed circle lies within twice the solid one.
    At 30~mrad $2\alpha/\lambda g_{111} = 2.9$ and the disks overlap strongly, whereas at 8~mrad $2\alpha/\lambda g_{111} = 0.8$ and no reflection overlaps the direct beam.
  }
  \label{fig:ptycho_60kV}
\end{figure*}

\begin{figure*}[ht]
\centering
\includegraphics[width = \textwidth]{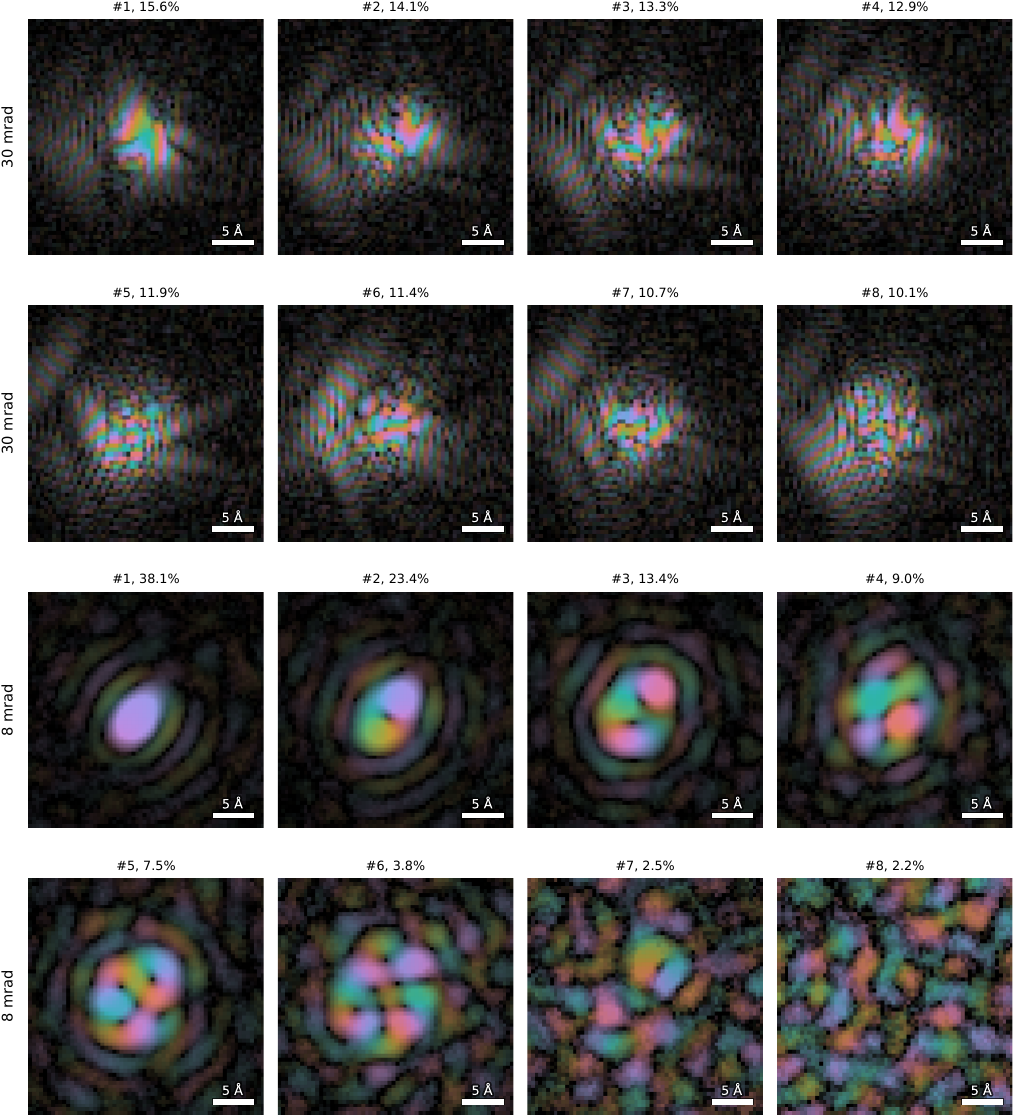}
  \caption{
    Real-space reconstructed probes from the 30 and 8~mrad reconstructions in~\cref{fig:ptycho_60kV} with associated mode weights.
  }
  \label{fig:si_real_probes_60kV_pair}
\end{figure*}

\begin{figure*}[ht]
\centering
\includegraphics[width = \textwidth]{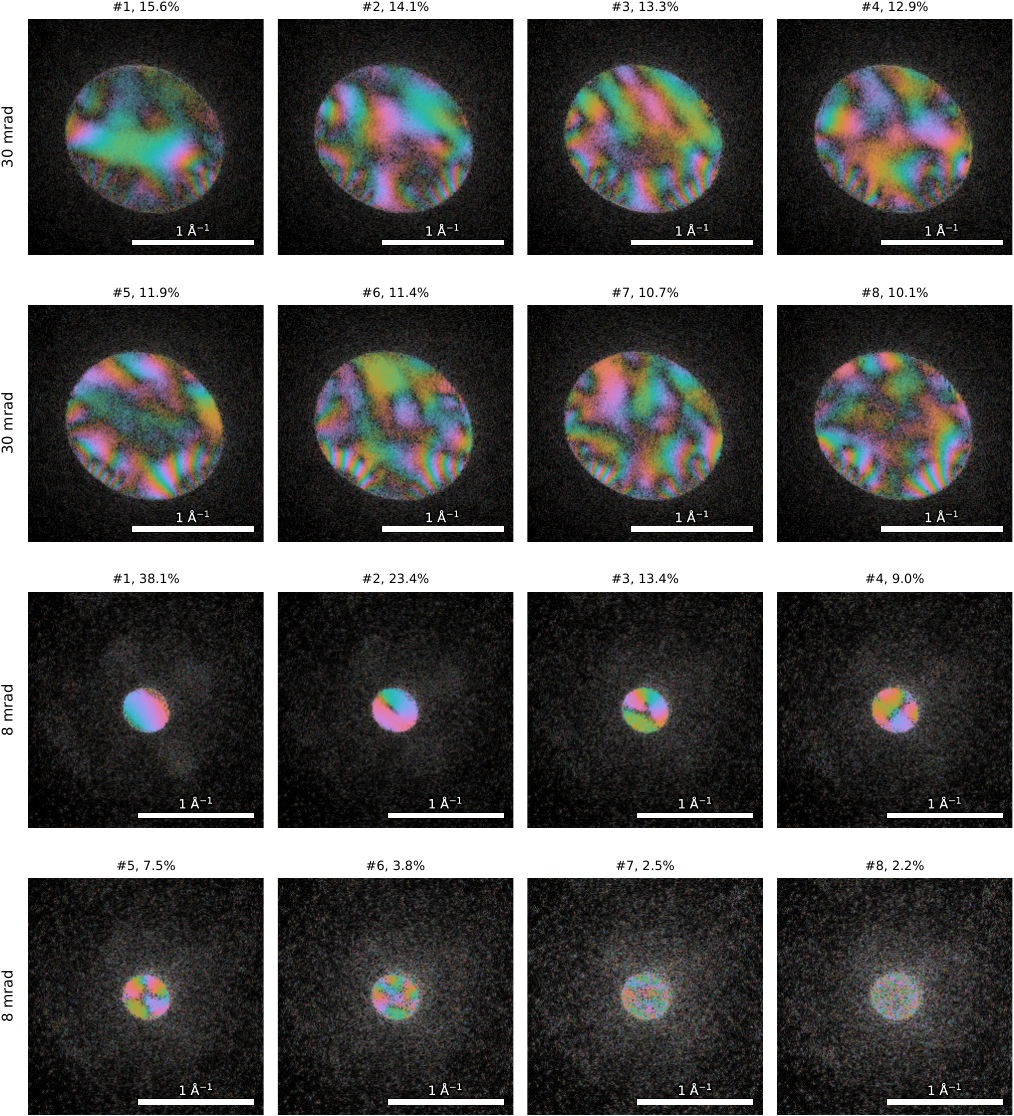}
  \caption{
    Reciprocal-space reconstructed probes from the 30 and 8~mrad reconstructions in~\cref{fig:ptycho_60kV} with associated mode weights.
  }
  \label{fig:si_reciprocal_probes_60kV_pair}
\end{figure*}

\begin{figure*}[ht]
\centering
\includegraphics[width = \textwidth]{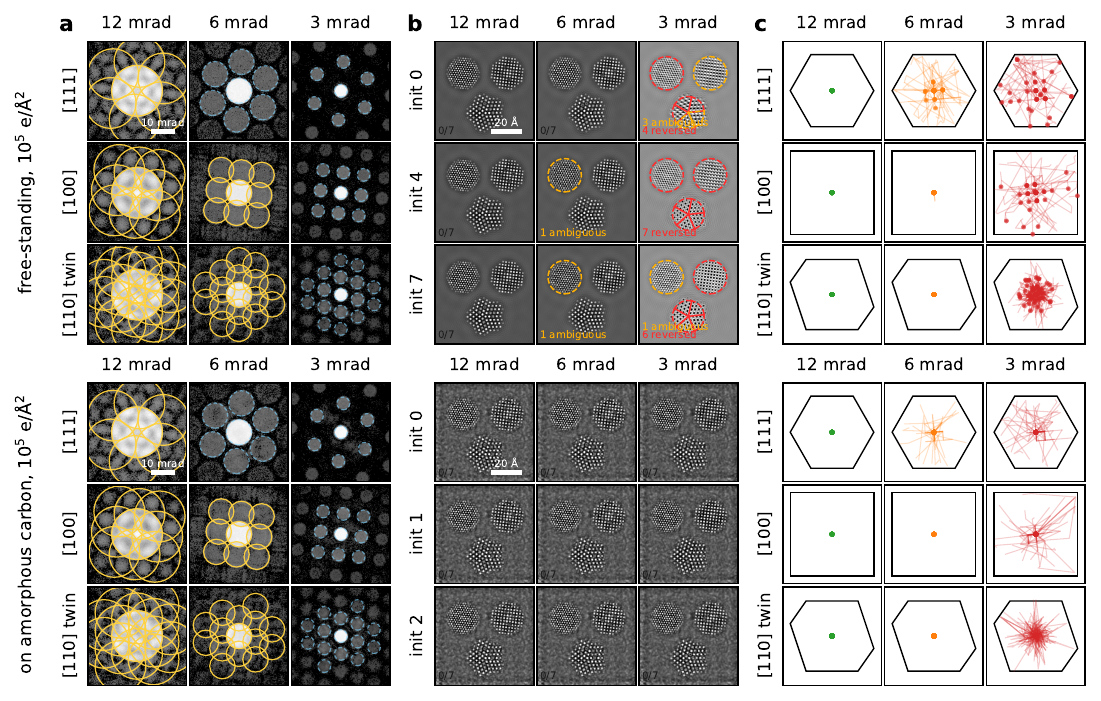}
  \caption{
    The amorphous substrate is what makes the reconstruction unique.
    Two rows at $10^5$~\eA{} with a matched single-slice forward model, differing only in whether the particles rest on amorphous carbon.
    Free-standing (top), the reconstruction is non-unique in every (draw, region) group at 3~mrad, with 44\% of regions reversed. This holds at every dose tested including noiseless, establishing a structural limit rather than a counting-statistics one.
    On carbon (bottom) the same measurement is unanimous, 0.00, at $10^5$~\eA{} and above.
    The mechanism is visible in (a): the free-standing patterns show isolated disks on an empty background, while the supported patterns show diffuse scattering filling the gaps between them, giving the Bragg beams the support in the inter-disk gaps described by~\cref{eq:product}.
    Panel (c) summarizes the consequence.
    Offsets fill the Wigner--Seitz cell without the substrate, and collapse to a point with it.
    Panels (a)--(c) follow~\cref{fig:sweep}.
    }
  \label{fig:si_substrate}
\end{figure*}

\begin{figure*}[ht]
\centering
\includegraphics[width = \textwidth]{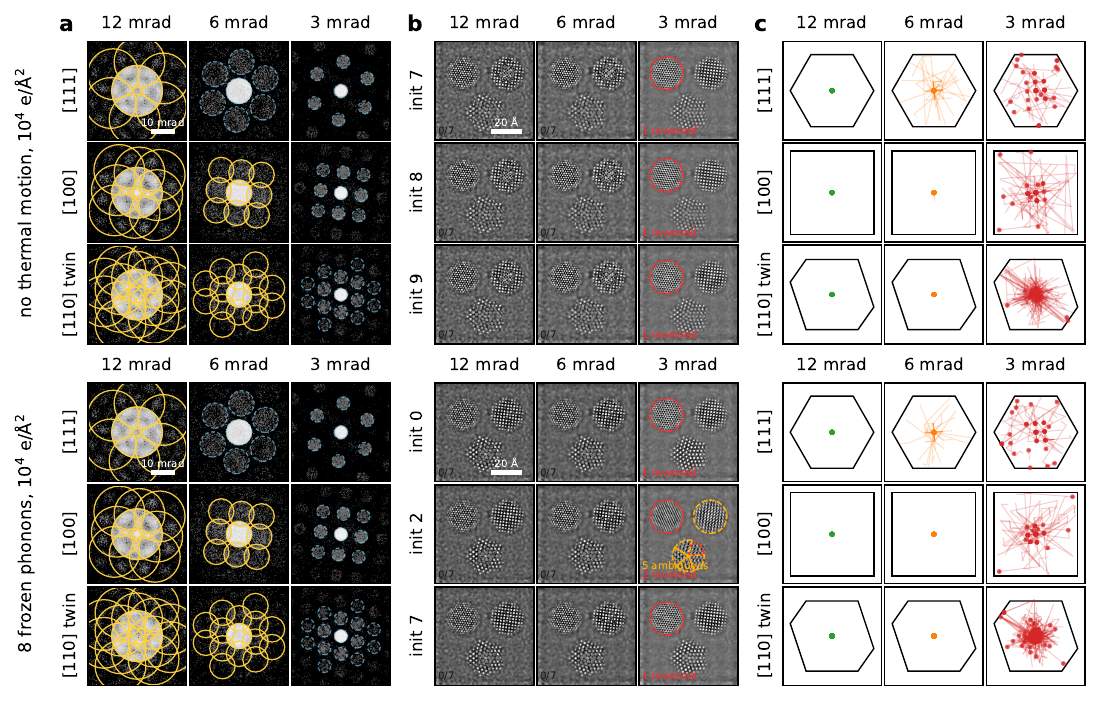}
  \caption{
    Thermal motion degrades the reconstruction and makes it ambiguous at the same time, and that combination is the hardest case to diagnose.
    Two rows of the same multislice data at $10^4$~\eA, differing only in whether the forward model was averaged over eight frozen-phonon configurations.
    Phonons raise 3~mrad non-uniqueness from 0.21 to 0.32, but unlike the pure phase object mismatch of~\cref{fig:sweep} they do not leave reconstruction quality intact:
    mean $|\mathrm{corr}|$ falls from 0.793 to 0.696 and the degraded fraction triples from 0.05 to 0.16.
    The two failures are visible together in (b), where \texttt{init 2} returns five ambiguous and two reversed regions in a visibly grainier reconstruction.
    Panels (a)--(c) follow~\cref{fig:sweep}.
  }
  \label{fig:si_phonons}
\end{figure*}

\begin{figure*}[ht]
\centering
\includegraphics[width = \textwidth]{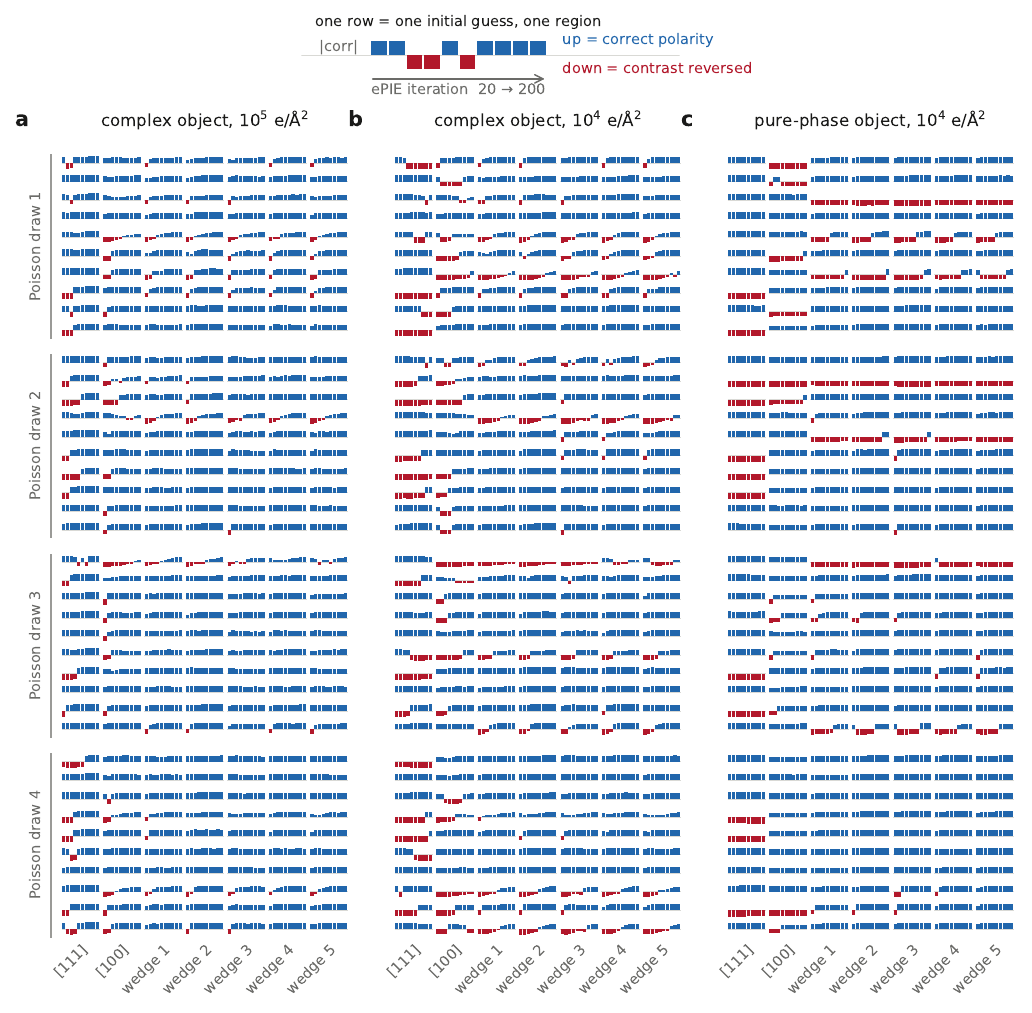}
  \caption{
    Every reconstruction behind~\cref{fig:sweep}, at 3~mrad, resolved by initial guess.
    Each row is one initial guess on one region, and the bars trace ePIE iterations 20 to 200.
    Bar height is $|\mathrm{corr}|$ against the band-limited ground truth, and bars are drawn upwards in blue where the polarity is correct and downwards in red where the contrast is reversed.
    Rows are grouped into the four Poisson draws, ten initial guesses each, and columns are the seven analyzed regions.
    Were the reconstruction uniquely determined, every row within a draw would be identical.
    (a) A complex-valued object at $10^5$~\eA{} is essentially all blue, with only brief early excursions that resolve well before iteration 200.
    (b) Dropping the dose to $10^4$~\eA{} leaves whole rows reversed, and the polarity of a given region now depends on the initial guess.
    (c) Constraining the object to pure phase reverses entire blocks, and the fourth Poisson draw is visibly cleaner than the first three, which is the draw-to-draw spread quoted in the text.
    Note throughout that bar heights barely change, whether a run is blue or red.
    The lattice, its spacing and its orientation are recovered correctly in every case, and it is only the polarity and the origin that scatter.
  }
  \label{fig:si_sparklines}
\end{figure*}

\begin{table*}[t]
\centering
\caption{
  Simulation results for every condition tested. The three multislice conditions were swept at 3~mrad only, since 6 and 12~mrad are unanimous throughout for the matched model.
  Non-uniqueness, reversed fraction, shifted fraction, mean $|\mathrm{corr}|$ and degraded fraction for every condition tested, as defined in~\cref{sec:methods_metrics}.
  The reversed and shifted columns separate the two gauge freedoms, and can be read against each other since both are rates over the same regions.
  The runs column gives the number of reconstructions behind each row and groups the number of (draw, region) pairs over which non-uniqueness is evaluated.
  Noisy conditions comprise 4 Poisson draws $\times$ 10 initial guesses over 7 regions; noiseless conditions are deterministic, so a single dataset exists and they carry one block of 10 initial guesses over the same 7 regions.
}
\label{tab:si_summary}
\small
\begin{tabular}{llrrrrrrrr}
\hline
condition & angle & dose & runs & groups & non-unique & reversed & shifted & $|\mathrm{corr}|$ & degraded \\
\hline
matched, on carbon & 12~mrad & noiseless & 50 & 7 & 0.00 & 0.00 & 0.00 & 0.958 & 0.00 \\
matched, on carbon & 12~mrad & $10^{6}$ & 40 & 28 & 0.00 & 0.00 & 0.00 & 0.947 & 0.00 \\
matched, on carbon & 12~mrad & $10^{5}$ & 40 & 28 & 0.00 & 0.00 & 0.00 & 0.916 & 0.00 \\
matched, on carbon & 12~mrad & $10^{4}$ & 40 & 28 & 0.00 & 0.00 & 0.00 & 0.893 & 0.00 \\
matched, on carbon & 6~mrad & noiseless & 50 & 7 & 0.00 & 0.00 & 0.00 & 0.957 & 0.00 \\
matched, on carbon & 6~mrad & $10^{6}$ & 40 & 28 & 0.00 & 0.00 & 0.00 & 0.949 & 0.00 \\
matched, on carbon & 6~mrad & $10^{5}$ & 40 & 28 & 0.00 & 0.00 & 0.00 & 0.915 & 0.00 \\
matched, on carbon & 6~mrad & $10^{4}$ & 40 & 28 & 0.00 & 0.00 & 0.00 & 0.890 & 0.00 \\
matched, on carbon & 3~mrad & noiseless & 50 & 7 & 0.00 & 0.00 & 0.00 & 0.900 & 0.00 \\
matched, on carbon & 3~mrad & $10^{6}$ & 40 & 28 & 0.00 & 0.00 & 0.00 & 0.900 & 0.00 \\
matched, on carbon & 3~mrad & $10^{5}$ & 40 & 28 & 0.00 & 0.00 & 0.00 & 0.897 & 0.00 \\
matched, on carbon & 3~mrad & $10^{4}$ & 40 & 28 & 0.32 & 0.05 & 0.05 & 0.808 & 0.02 \\
\hline
matched, free-standing & 12~mrad & noiseless & 10 & 7 & 0.00 & 0.00 & 0.00 & 0.992 & 0.00 \\
matched, free-standing & 12~mrad & $10^{6}$ & 40 & 28 & 0.00 & 0.00 & 0.00 & 0.980 & 0.00 \\
matched, free-standing & 12~mrad & $10^{5}$ & 40 & 28 & 0.00 & 0.00 & 0.00 & 0.939 & 0.00 \\
matched, free-standing & 12~mrad & $10^{4}$ & 40 & 28 & 0.00 & 0.00 & 0.00 & 0.908 & 0.00 \\
matched, free-standing & 6~mrad & noiseless & 10 & 7 & 0.00 & 0.00 & 0.00 & 0.966 & 0.00 \\
matched, free-standing & 6~mrad & $10^{6}$ & 40 & 28 & 0.00 & 0.00 & 0.00 & 0.964 & 0.00 \\
matched, free-standing & 6~mrad & $10^{5}$ & 40 & 28 & 0.07 & 0.00 & 0.00 & 0.931 & 0.00 \\
matched, free-standing & 6~mrad & $10^{4}$ & 40 & 28 & 0.04 & 0.00 & 0.00 & 0.893 & 0.00 \\
matched, free-standing & 3~mrad & noiseless & 10 & 7 & 1.00 & 0.45 & 0.03 & 0.748 & 0.01 \\
matched, free-standing & 3~mrad & $10^{6}$ & 40 & 28 & 1.00 & 0.37 & 0.03 & 0.806 & 0.03 \\
matched, free-standing & 3~mrad & $10^{5}$ & 40 & 28 & 1.00 & 0.44 & 0.09 & 0.816 & 0.01 \\
matched, free-standing & 3~mrad & $10^{4}$ & 40 & 28 & 1.00 & 0.45 & 0.17 & 0.829 & 0.00 \\
\hline
multislice, complex & 3~mrad & noiseless & 10 & 7 & 0.00 & 0.00 & 0.00 & 0.824 & 0.00 \\
multislice, complex & 3~mrad & $10^{6}$ & 40 & 28 & 0.00 & 0.00 & 0.00 & 0.832 & 0.00 \\
multislice, complex & 3~mrad & $10^{5}$ & 40 & 28 & 0.00 & 0.00 & 0.00 & 0.821 & 0.01 \\
multislice, complex & 3~mrad & $10^{4}$ & 40 & 28 & 0.21 & 0.05 & 0.06 & 0.793 & 0.05 \\
\hline
multislice, pure-phase & 3~mrad & noiseless & 10 & 7 & 0.71 & 0.06 & 0.09 & 0.804 & 0.00 \\
multislice, pure-phase & 3~mrad & $10^{6}$ & 40 & 28 & 0.21 & 0.05 & 0.05 & 0.816 & 0.00 \\
multislice, pure-phase & 3~mrad & $10^{5}$ & 40 & 28 & 0.25 & 0.05 & 0.06 & 0.842 & 0.00 \\
multislice, pure-phase & 3~mrad & $10^{4}$ & 40 & 28 & 0.82 & 0.12 & 0.19 & 0.818 & 0.00 \\
\hline
multislice + phonons & 3~mrad & noiseless & 10 & 7 & 0.00 & 0.00 & 0.00 & 0.810 & 0.01 \\
multislice + phonons & 3~mrad & $10^{6}$ & 40 & 28 & 0.00 & 0.00 & 0.00 & 0.808 & 0.04 \\
multislice + phonons & 3~mrad & $10^{5}$ & 40 & 28 & 0.00 & 0.00 & 0.00 & 0.791 & 0.04 \\
multislice + phonons & 3~mrad & $10^{4}$ & 40 & 28 & 0.32 & 0.06 & 0.08 & 0.696 & 0.16 \\
\hline
\end{tabular}
\end{table*}

\end{document}